\documentclass[twocolumn,aps,prc,superscriptaddress,showpacs,floatfix,longbibliography]{revtex4-1}
\usepackage{url}
\usepackage{cancel}
\usepackage[colorlinks,linkcolor=blue,citecolor=blue,filecolor=black,urlcolor=blue]{hyperref}
\usepackage{epsfig,graphics}
\usepackage{graphicx}
\usepackage{dcolumn}
\usepackage{bm}
\usepackage[usenames]{color}
\usepackage{amssymb}
\usepackage{amsmath}
\usepackage{multirow}
\usepackage{float}
\usepackage{harpoon}
\usepackage{MnSymbol}
\usepackage{appendix}
\usepackage{color}
\usepackage{hyperref}
\usepackage{cleveref}

\begin{document}

\title{Shear viscosity of nuclear matter in the spinodal region}
\author{Lei-Ming Hua}
\affiliation{Shanghai Institute of Applied Physics, Chinese Academy of Sciences, Shanghai 201800, China}
\affiliation{University of Chinese Academy of Sciences, Beijing 100049, China}
\author{Jun Xu}\email[Correspond to\ ]{xujun@zjlab.org.cn}
\affiliation{School of Physics Science and Engineering, Tongji University, Shanghai 200092, China}
\affiliation{Shanghai Advanced Research Institute, Chinese Academy of Sciences, Shanghai 201210, China}
\affiliation{Shanghai Institute of Applied Physics, Chinese Academy of Sciences, Shanghai 201800, China}

\begin{abstract}
Based on IBUU simulations calibrated by previous efforts of the transport model evaluation project, we have studied the specific shear viscosity $\eta/s$ of nuclear matter in the spinodal region using the Green-Kubo method. With the momentum-independent mean-field potential which reproduces reasonably well empirical nuclear matter properties and nuclear phase diagram, we have generated dynamically stable and thermalized nuclear cluster systems in a box with the periodic boundary condition. Extensive results of the $\eta/s$ at different average densities and temperatures in uniform and non-uniform systems are compared, and we found that the shear viscosity is smaller with nuclear clusters due to the enhanced correlation of the energy-momentum tensor and the stronger collision effect. The temperature dependence of the $\eta/s$ has a minimum only at low average densities of $\rho<0.3\rho_0$. The present study serves as a rigorous baseline calculation of the $\eta/s$ in nuclear systems with clusters, and helps to understand the relation between the shear viscosity and the nuclear phase diagram.
\end{abstract}
\maketitle

\section{Introduction}
\label{sec:intro}

Transport properties of nuclear matter is important for understanding the dynamics in intermediate-energy heavy-ion collisions and the behavior of the nuclear liquid-gas phase transition. In the past decades, the shear viscosity of strong interacting matter has been studied with various approaches. By comparing collective flows from hydrodynamic simulations with the experimental data, it was found that quark-gluon plasma produced in ultrarelativistic heavy-ion collisions is a nearly ideal fluid and has a very small specific shear viscosity~\cite{Peshier:2005pp,Majumder:2007zh,Song:2010mg,Schenke:2010rr,Bernhard:2019bmu,Parkkila:2021yha}, i.e., the ratio $\eta/s$ of the shear viscosity $\eta$ to the entropy density $s$ is only a few times of the Kovtun-Son-Starinets (KSS) bound~\cite{Kovtun:2004de}. In heavy-ion collisions at lower collision energies, where the dynamics is mostly dominated by hadron resonance gas or nucleon degree of freedom, the specific shear viscosity is much larger (see, e.g., Ref.~\cite{Reichert:2020oes}). Interestingly, a minimum value of $\eta/s$ is observed around the temperature of the hadron-quark phase transition~\cite{Csernai:2006zz,Lacey:2006bc}. In the presence of the liquid-gas phase transition in nuclear matter, a minimum $\eta/s$ is also observed based on different approaches~\cite{Chen:2007xe,Pal:2010sj,Xu:2013nwa,Xu:2015lna,PhysRevC.105.064613}. In this sense, the behavior of $\eta/s$ is related to the phase diagram of the strong interacting matter (see, e.g., Refs.~\cite{Ghosh:2014vja,Grefa:2022sav}).

Among various approaches of studying the shear viscosity of hadron resonance gas~\cite{Muronga:2003tb,Chen:2006iga,Demir:2008tr,Rose:2017bjz} and nuclear matter~\cite{Danielewicz:1984kt,Shi:2003np}, directly using the Green-Kubo formula~\cite{Kubo_1966} is the most rigorous way of the study (see Ref.~\cite{Plumari:2012ep} for the comparison of different approaches). On the other hand, the accurate calculations of the shear viscosity by using the Green-Kubo formula requires a well calibrated transport model~\cite{Xu:2019hqg}, and such simulation is generally carried out in a box system with the periodic boundary condition~\cite{Motornenko:2017hob,Deng:2021rpq}. In the semiclassical approximation, the shear viscosity is proportional to $1/\sigma$, with $\sigma$ being the scattering cross section, so reproducing the theoretical limit of the collision rate is crucial for obtaining an accurate shear viscosity via the Green-Kubo formula. Fortunately, this has bee achieved in Ref.~\cite{Zhang:2017esm} where different collision treatments were compared in detail and a few optimized collision treatments, which are necessary for reproducing the Boltzmann limit of the collision rate, were recommended. In order to study the behavior of the shear viscosity in the presence of the nuclear liquid-gas phase transition, a well calibrated mean-field calculation is needed to generate reasonable density fluctuations, and this has also been achieved in Ref.~\cite{Colonna:2021xuh} by comparing the resulting response function of the density fluctuation with the theoretical limit  predicted by the Landau parameter of the mean-field interaction.

In the present study, we investigate the behavior of the specific shear viscosity in the spinodal region of isospin symmetric nuclear matter based on a well calibrated isospin-dependent Boltzmann-Uehling-Uhlenbeck (IBUU) transport model, where nucleon-nucleon elastic scatterings are implemented by using a modified Bertsch's prescription~\cite{Bertsch:1988ik}, and the mean-field evolution is simulated by using a lattice Hamiltonian framework~\cite{Lenk:1989zz}. The phase diagram of nuclear matter is obtained from a simplified momentum-independent potential, which reproduces empirical nuclear matter properties at the saturation density. This mean-field potential is also implemented in the dynamical simulation to generate density fluctuations in the spinodal region. After the density evolution reaches a dynamic equilibrium, the Green-Kubo formula, for which we will show to be also valid in non-uniform systems, is then used to calculate the shear viscosity from the correlation of the energy-momentum tensor. Results from uniform and non-uniform nuclear systems are compared, and we found that the density fluctuations due to the nuclear liquid-gas phase transition reduce considerably both $\eta$ and $\eta/s$.

The rest part of the paper is organized as follows. Section~\ref{sec:theory} gives the theoretical framework, including thermodynamic properties of isospin symmetric nuclear matter, transport simulations in a box system, and the Green-Kubo method for calculating the shear viscosity. Section~\ref{sec:results} presents the way to generate dynamic and thermal equilibrated clusterizations from transport simulations in a box system, and discusses the corresponding behavior of the specific shear viscosity from the Green-Kubo method in the nuclear liquid-gas mixed phase. We conclude and outlook in Sec.~\ref{sec:summary}.

\section{Theoretical framework}
\label{sec:theory}

With a simple nuclear mean-field potential that reproduces empirical nuclear matter properties around the saturation density, we briefly present in this section the main features of the thermodynamics and the phase diagram of isospin symmetric nuclear matter. Details on transport simulations in a box system with the periodic boundary condition will also be provided, and the major focuses will be on the treatments of nucleon-nucleon elastic collisions with a modified Bertsch's prescription and the mean-field evolution based on the lattice Hamiltonian framework. We will further discuss how we obtain the shear viscosity through the Green-Kubo method.

\subsection{Thermodynamics of nuclear matter}

We adopt in the present study the following momentum-independent single-nucleon potential in isospin symmetric nuclear matter of density $\rho$ as
\begin{equation}
\label{eq10}
U(\rho)=\alpha \left(\frac{\rho}{\rho_{0}}\right)+\beta\left(\frac{\rho}{\rho_{0}}\right)^{\gamma},
\end{equation}
with coefficients $\alpha = -0.218$ GeV, $\beta = 0.164$ GeV, and $\gamma = 4/3$ which reproduce the saturation density $\rho_0=0.16$ fm$^{-3}$, the binding energy $E_0=-16$ MeV at $\rho_0$, and the incompressiblity $K_0=237$ MeV. Although the above mean-field potential is simple and far from realistic, we will see that it reproduces the main features of the nuclear phase diagram and is adequate for the present study.

With the single-nucleon potential given above, the corresponding potential energy density $\epsilon_p$ is then written as
\begin{equation}
\label{eq11}
\epsilon_p=\frac{\alpha}{2} \frac{\rho^{2}}{\rho_{0}}+\frac{\beta}{1+\gamma} \frac{\rho^{1+\gamma}}{\rho_{0}^{\gamma}}.
\end{equation}
In the quasi-free nucleon approximation, the kinetic energy density $\epsilon_k$ can be expressed as
\begin{equation}
\label{ek}
\epsilon_k=4\int \frac{d^{3} p}{(2\pi)^3} \left(\sqrt{p^2+m^2}-m\right) f(\vec{r}, \vec{p}) ,
\end{equation}
where $m=939$ MeV is the bare nucleon mass, $f(\vec{r}, \vec{p})$ is the nucleon phase-space distribution function, which in the thermal equilibrated system at temperature $T$ is the Fermi-Dirac distribution expressed as
\begin{equation}
\label{eq20}
f(\vec{r}, \vec{p})=\frac{1}{\exp \left(\frac{\sqrt{p^2+m^2}-m+U-\mu}{T}\right)+1}.
\end{equation}
In the above, $\mu$ is the nucleon chemical potential determined by
\begin{equation}
\label{eq12}
\rho=4 \int \frac{d^{3} p}{(2\pi)^3}  f(\vec{r}, \vec{p}) .
\end{equation}
For a quasi-free Fermion system, the entropy density can be calculated from the phase-space distribution function through the expression
\begin{equation}\label{s}
s = - 4 \int \frac{d^{3} p}{(2\pi)^3} [f \ln f+(1-f) \ln (1-f)].
\end{equation}

The above relations give densities of quantities at local position $\vec{r}$, while for a uniform system the binding energy per nucleon can be expressed as $E=\epsilon/\rho$ with $\epsilon=\epsilon_p+\epsilon_k$ being the total energy density, and the pressure $P$ can be calculated from the thermodynamic relation
\begin{equation}
\label{eq23}
P=Ts-\epsilon+\mu\rho.
\end{equation}
The pressure can be used to identify the spinodal region of the nuclear matter (see, e.g., Ref.~\cite{Xu:2007eq}), corresponding to a liquid-gas mixed phase~\cite{Muller:1995ji,Chomaz:2003dz}. In the region of
\begin{equation}
\label{eq24}
\left(\frac{\partial P}{\partial \rho}\right)_{T} < 0,
\end{equation}
the system is mechanically unstable. This is because increasing (reducing) the local density reduces (increases) the local pressure so more particles will flow into (away from) the local area, further reducing (increasing) the local pressure, thus any small density fluctuations may grow and the nuclear matter becomes mechanically unstable. We discuss properties of isospin symmetric nuclear matter in the present study, and neglect the chemical instability which exists only in isospin asymmetric nuclear matter.

\begin{figure}[!h]
\includegraphics[width=1.0\linewidth]{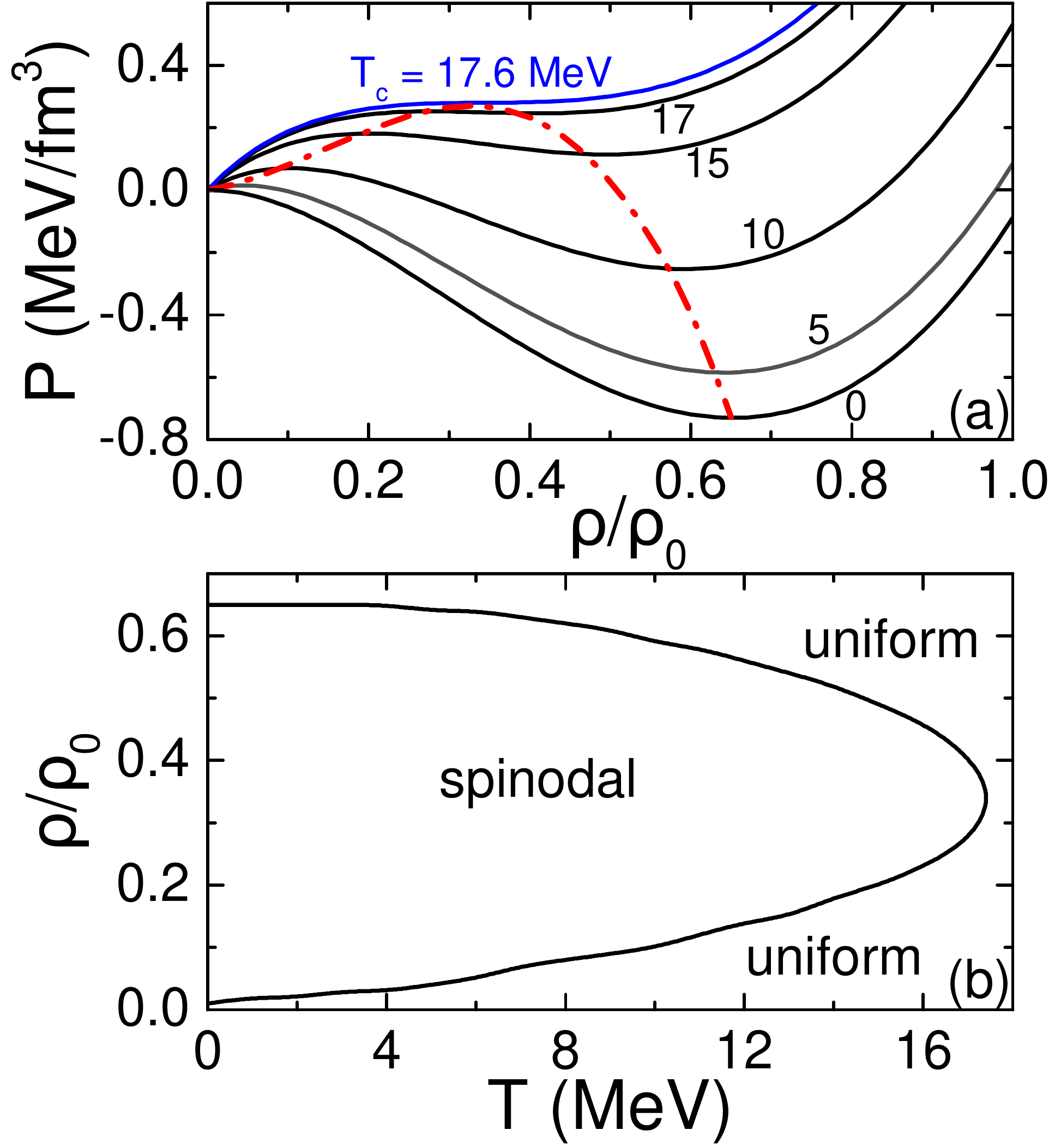}
\caption{\label{prhoT} Upper: Pressure of hot nuclear matter at different temperatures as a function of nucleon density $\rho$; Lower: Boundaries of mechanical instability in the $(\rho, T)$ plane. }
\end{figure}

Figure~\ref{prhoT} (a) displays the pressure of nuclear matter at different temperatures $T$ based on the nucleon mean-field potential as Eq.~(\ref{eq10}). The mechanical instability region that satisfies Eg.~(\ref{eq24}) shrinks with increasing temperature, and disappears at $T_c=17.6$ MeV, above which the pressure $P$ increases monotonically with increasing density $\rho$. The boundary of the mechanical instability region is determined by
\begin{equation}
\label{eq25}
\left(\frac{\partial P}{\partial \rho}\right)_{T}=0.
\end{equation}
A simple phase diagram of isospin symmetric nuclear matter is plotted in Fig.~\ref{prhoT} (b) in the $(\rho, T)$ plane. As mentioned above, density fluctuations are expected to appear within the spinodal region, where the liquid phase with a higher density and the gas phase with a lower density coexist. Outside the spinodal region, the nuclear matter is expected to stay uniform.

\subsection{Transport simulation in a box system}

The simulation is carried out in a cubic box with a length $L = 20$ fm in x, y, and z directions based on the IBUU transport model. The periodic boundary condition is applied, i.e., a nucleon that leaves the system on one side of the cubic box will enter the box from another side with the same momentum, and the distance between two nucleons in each dimension is less than $L/2$ considering the periodic nature of the system. We use $1000$ test particles for each nucleon, to assure accurate calculations of the mean-field evolution and the Pauli blocking, and we find that the results remain almost unchanged by further increasing the number of test particles. The initial coordinates of nucleons are uniformly distributed within the box, and the initial momenta of nucleons are sampled according to the Fermi-Dirac distribution for a given temperature and density. A time step $\Delta t = 0.5$ fm/c is used for both nucleon-nucleon collisions and the nucleon propagation under the mean-field potential.

For elastic nucleon-nucleon collisions, we use the geometric method as in the appendix B of Ref.~\cite{Bertsch:1988ik} but with some modifications. In Bertsch's prescription, the minimum distance of two colliding particles in their center-of-mass (C.M.) frame perpendicular to their relative velocity is
\begin{equation}
{d_\perp^\star}^2 = (\vec{r}_1^\star-\vec{r}_2^\star)^2 - \frac{[(\vec{r}_1^\star-\vec{r}_2^\star)\cdot \vec{v}_{12}^\star]^2}{{v_{12}^\star}^2},
\end{equation}
where $\vec{r}_1^\star$ and $\vec{r}_2^\star$ are positions of the two particles, and $\vec{v}_{12}^\star = \vec{v}_1^\star - \vec{v}_2^\star$ is their relative velocity, with the asterisk representing the quantity in the C.M. frame of the colliding particles. The collision can happen if the condition
\begin{equation}
\pi {d_\perp^\star}^2 < \sigma
\end{equation}
is satisfied, and we use a constant and isotropic nucleon-nucleon cross section $\sigma=40$ mb in the present study. Whether the collision happens in this time step is determined by the condition of the closest approach, i.e.,
\begin{equation}
|(\vec{r}_1^\star-\vec{r}_2^\star)\cdot \vec{v}_{12}^\star/{v_{12}^\star}^2|< \frac{1}{2} \delta t.
\end{equation}
The relation between $\delta t$ and $\Delta t$ has not been specified in Ref.~\cite{Bertsch:1988ik}. We set $\delta t = \Delta t/\gamma$, where $\gamma=1/\sqrt{1-\beta^2}$ is the Lorentz factor with $\beta$ being the average velocity of the colliding pair in the box frame.

With the original Bertsch's prescription for collisions, the particle pair that collide once have $50\%$ chance to collide again in the subsequent time step for an isotropic cross section, if the final velocities point toward each other. This effect is contradictory to the assumption of the Boltzmann equation that the collisions are independent of each other and are not repeated. These spurious collisions can be avoided by setting that the two particles, that have collided once, can not collide again unless one of them has collided with a third particle. By doing this, we remove the leading-order correlations induced by collisions, while higher-order correlations still remain and can affect the collision rate especially at high densities or with a large nucleon-nucleon cross section. For more details about the collision criterion, we refer the reader to Refs.~\cite{Zhang:2017esm,Xu:2019hqg}. For the systems with the density below $\rho_0$ and the temperature of about $T \sim 10$ MeV, and especially with Pauli blocking, we will show that the above collision treatment is good enough to achieve the attempted collision rate from the theoretical limit.

Due to the Fermionic nature of nucleons, the collision can happen only if the final state of either colliding nucleon has not been occupied. The Pauli blocking probability is $1-(1-f_1)(1-f_2)$, where $f_1$ and $f_2$ are the local phase-space distribution functions for the final states of the colliding nucleon 1 and 2. To obtain the local phase-space distribution function, we divide the box system into cells and assume that the local thermal equilibrium is always maintained in each cell of the volume $2\times2\times2$ fm$^{3}$. The local phase-space distribution is calculated according to Eq.~(\ref{eq20}), where the temperature $T$ and the chemical potential $\mu$ are determined from simulations.

Although point particles are used for collisions and Pauli blocking, the mean-field evolution is based on the lattice Hamiltonian framework using finite-size test particles. The coordinate space is divided into cubic cells with the volume $l^3$, and the density at the site $\vec{r}_\alpha$ of the lattice is then given by
\begin{equation}\label{rhol}
\rho_L (\vec{r}_\alpha) = \frac{1}{N_{TP}}\sum_{i=1}^{A N_{TP}} G(\vec{r}_\alpha-\vec{r}_i),
\end{equation}
where $N_{TP}$ is the test-particle number per nucleon, $A$ is the total nucleon number determined by the average density, and $G$ is the shape function defined as
\begin{equation}\label{shape}
G(\vec{r}_\alpha-\vec{r}_i) = \frac{1}{(nl)^6}g(x)g(y)g(z)
\end{equation}
with $x=x_\alpha-x_i$, $y=y_\alpha-y_i$, $z=z_\alpha-z_i$, and
\begin{equation}
g(q)=(nl-|q|)\Theta(nl-|q|),
\end{equation}
where $l$ is the lattice spacing, $n$ determines the range of $G$, and $\Theta$ is the Heaviside function. We set $n=2$ and $l=1$ fm in the present study. The total potential energy of the system is the sum of that in each cubic cell, i.e., $E_p=l^3 \sum_\alpha \epsilon_p[\rho_L(\vec{r}_\alpha)]$. For a momentum-independent mean-field potential $U$ as in the present study, the canonical equations of motion for the $i$th nucleon can be expressed as
\begin{eqnarray}
\frac{d\vec{r}_i}{dt} &=& \frac{\vec{p}_i}{\sqrt{\vec{p}_i^2+m^2}},\\
\frac{d\vec{p}_i}{dt} &=& -l^3 \sum_\alpha \frac{\partial \epsilon_p[\rho_L(\vec{r}_\alpha)]}{\partial \rho_L} \frac{\partial \rho_L}{\partial \vec{r}_i} \notag\\
&=& -\frac{l^3}{N_{TP}} \sum_\alpha U[\rho_L(\vec{r}_\alpha)] \frac{\partial G(\vec{r}_\alpha-\vec{r}_i)}{\partial \vec{r}_i}.\label{eom}
\end{eqnarray}
We note that the relativistic kinematics is used, consistent with Eqs.~(\ref{ek}) and (\ref{eq20}) as well as the collision treatment. The accurate mean-field evolution with rigorous energy conservation can be achieved by solving numerically the above differential equations. For more details, we refer the reader to Refs.~\cite{Lenk:1989zz,Colonna:2021xuh}.

In the default calculation of the present study, we don't include the Coulomb potential, which is excepted to have no effect in uniform systems but may have some influence in non-uniform systems. In the lattice Hamiltonian framework, the Coulomb force acting on the $i$th particle is calculated from the Coulomb potential energy density $V^{cou}_\alpha$ according to
\begin{equation}\label{coulombf}
\left(\frac{d\vec{p}_i}{dt}\right)_c = \vec{F}_c = -l^3 Z_i e^2 \sum_\alpha \frac{\partial V^{cou}_\alpha(\vec{r}_\alpha)}{\partial \vec{r}_i},
\end{equation}
with $Z_i$ being the charge number of the $i$th nucleon, and the summation over the lattice sites $\vec{r}_\alpha$. Including both the direct and exchange contributions from the Coulomb interaction, the Coulomb potential energy density $V^{cou}_\alpha$ can be expressed as
\begin{equation}\label{coulombv}
V^{cou}_\alpha = \frac{l^3}{2} \sum_{\alpha',\alpha' \neq \alpha} \frac{\rho_L^c(\vec{r}_\alpha) \rho_L^c(\vec{r}_{\alpha'}) }{|\vec{r}_\alpha - \vec{r}_{\alpha'}|}
- \frac{3}{4} \left[ \frac{3 \rho_L^c (\vec{r}_\alpha)}{\pi} \right]^{4/3}.
\end{equation}
In the above, $\rho_L^c(\vec{r}_\alpha)$ is the net-charge number density at the lattice site $\vec{r}_\alpha$ and is calculated in a way similar to Eq.~(\ref{rhol}) for the particle density.  Substituting Eq.~(\ref{coulombv}) into Eq.~(\ref{coulombf}) leads to the following Coulomb force acting on the $i$th charged particle
\begin{eqnarray}\label{coulat}
\vec{F}_c &=& - \frac{l^3 Z_i e^2}{N_{TP}} \sum_\alpha \left\{ \frac{l^3}{N_{TP}}\sum_{\alpha',\alpha' \neq \alpha} \frac{N_{TP}\rho_L^c(\vec{r}_\alpha)}{|\vec{r}_\alpha - \vec{r}_{\alpha'}|} \frac{\partial G(\vec{r}_{\alpha'} - \vec{r}_i)}{\partial \vec{r}_i}\right. \notag\\
&-& \left.\frac{3}{\pi} \left[ \frac{3 \rho_L^c (\vec{r}_\alpha)}{\pi}\right]^{1/3} \frac{\partial G(\vec{r}_\alpha - \vec{r}_i)}{\partial \vec{r}_i}
\right\}.
\end{eqnarray}

\subsection{Green-Kubo method}

The Green-Kubo formula relates linear transport coefficients to near-equilibrium correlations of dissipative fluxes, and treats dissipative fluxes as perturbations to local thermal equilibrium~\cite{Kubo_1966,Hosoya:1983id,Paech:2006st}. The shear viscosity from the Green-Kubo formula is expressed as
\begin{equation}
\label{eq2}
\eta=\frac{1}{T} \int d^{3} r \int_{t_0}^{\infty} d t\left\langle\pi^{x y}(\vec{0}, t_0) \pi^{x y}(\vec{r}, t)\right\rangle_{\text {equil }},
\end{equation}
where $T$ is the temperature of the system, $t-t_0$ is the post-equilibration time with $t_0$ being the time when the system has reached dynamic equilibrium, and $\pi^{x y} $ is the shear component of the energy-momentum tensor which can be expressed as
\begin{equation}
\label{eq3}
\pi^{x y}=\int \frac{d^{3} p}{(2\pi)^3} \frac{p^{x} p^{y}}{E} f(\vec{r}, \vec{p}),
\end{equation}
with $E=\sqrt{\vec{p}^2+m^2}$ being the nucleon energy. Given the momenta of test particles, the local $\pi^{x y}$ can be calculated from the summation of nucleons in a local cell
\begin{equation}
\label{eq4}
\pi^{x y}=\frac{1}{V_c} \sum_{i} \frac{p^{x}_i p^{y}_i}{E_i},
\end{equation}
where $V_c$ is the volume of the cell, and $p^{x}_i$, $p^{y}_i$, and $E_i=\sqrt{m^2+\vec{p}_i^2}$ are, respectively, the momentum in the $x$ and $y$ direction and the energy of the $i$th nucleon in the local cell obtained from transport simulations. The average over parallel events is applied in the calculation but omitted in the formulaes.

In the box system with the periodic boundary condition as in the present study, we show in the following that Eq.~(\ref{eq2}) can be calculated through
\begin{equation}
\label{eq2s}
\eta=\frac{V}{T} \int_{t_0}^{\infty} d t\left\langle\Pi^{x y}(t_0) \Pi^{x y}(t)\right\rangle_{\text {equil }},
\end{equation}
where $V=L^3$ is the volume of the box, and $\Pi^{x y}$ is calculated similar to Eq.~(\ref{eq4}) but by summing all nucleons in the box system. The interand in the above equation can be further expressed as
\begin{eqnarray}
&&\Pi^{x y}(t_0) \Pi^{x y}(t) \notag\\
&=&\frac{1}{V} \left(\sum_{i} \frac{p^{x}_i p^{y}_i}{E_i}\right)_{t_0} \frac{1}{V} \left(\sum_{j} \frac{p^{x}_j p^{y}_j}{E_j}\right)_t \notag\\
&=&\frac{1}{(N_cV_c)^2} \left(\sum_{c_1}\sum_{i_{c_1}} \frac{p^{x}_{i_{c_1}} p^{y}_{i_{c_1}}}{E_{i_{c_1}}}\right)_{t_0} \left(\sum_{c_2}\sum_{j_{c_2}} \frac{p^{x}_{j_{c_2}} p^{y}_{j_{c_2}}}{E_{j_{c_2}}}\right)_{t} \notag\\
&=&\frac{1}{(N_cV_c)^2} \sum_{c_1} \sum_{c_2} \left(\sum_{i_{c_1}} \frac{p^{x}_{i_{c_1}} p^{y}_{i_{c_1}}}{E_{i_{c_1}}}\right)_{t_0} \left(\sum_{j_{c_2}} \frac{p^{x}_{j_{c_2}} p^{y}_{j_{c_2}}}{E_{j_{c_2}}}\right)_{t}.
\end{eqnarray}
In the above, $N_c=V/V_c$ is the number of cells, and $\sum_{c_{1(2)}}$ in the third line represents the summation over all cells, with $i_{c_1}$ ($j_{c_2}$) being the nucleon label in cell $c_{1(2)}$. Comparing with Eq.~(\ref{eq2}) which should be independent of the choice for the cell $\vec{r}=0$ in a box with the periodic boundary condition, we can express $\pi^{x y}(\vec{0}, t_0)$ as
\begin{eqnarray}
\pi^{x y}(\vec{0}, t_0) = \frac{1}{N_c} \sum_{c_1} \left(\sum_{i_{c_1}} \frac{p^{x}_{i_{c_1}} p^{y}_{i_{c_1}}}{E_{i_{c_1}}}\right)_{t_0}.
\end{eqnarray}
The above relation can be understood since choosing different cell of $\vec{r}=0$ is identical to choosing different starting time $t_0$ or parallel events once the system has reached dynamic equilibrium. So one can now see that Eq.~(\ref{eq2s}) is identical to Eq.~(\ref{eq2}) and thus valid for both uniform and non-uniform systems.

\section{Results and discussions}
\label{sec:results}

Since the shear viscosity of nuclear matter is dominated by nucleon-nucleon collisions, we first compare the collision rate from IBUU simulations in a box system with the theoretical limit as in Ref.~\cite{Zhang:2017esm}. Afterwards, we discuss the way to prepare a dynamic and thermal equilibrated system in the spinodal region in the presence of both the nucleon mean-field potential and nucleon-nucleon collisions with Pauli blocking, for the calculation of the shear viscosity via the Green-Kubo method. Extensive density and temperature dependence of the specific shear viscosity will be investigated, and results of uniform and non-uniform systems will be compared.

\subsection{Calibrating the collision rate}

For a uniform box system, the theoretical limit of the collision rate, i.e., the total collision number of the system per unit time, can be expressed as
\begin{equation}\label{cr}
\frac{dN_{coll}}{dt} = \frac{1}{2}V\rho^2 \sigma \int d^{3} p_1  d^{3} p_2  v_{mol} \tilde f(p_1) \tilde f(p_2).
\end{equation}
In the above,
\begin{equation}
v_{mol} = \frac{\sqrt{(E_1 E_2 - \vec{p}_1 \cdot \vec{p}_2)^2 - m^4}}{E_1 E_2}
\end{equation}
is the M{\o}ller velocity with $E_{1(2)}=\sqrt{\vec{p}_{1(2)}^2+m^2}$. $\tilde f(p)$ is the normalized nucleon momentum distribution, which can be a Maxwell-Boltzmann (MB) distribution, i.e., $\tilde f(p)= \frac{1}{4\pi m^2 T K_2(m/T)} \exp\left(-\sqrt{p^2+m^2}/T\right)$, with $K_n$ being the $n$th-order modified Bessel function, or a Fermi-Dirac (FD) distribution similar to Eq.~(\ref{eq20}) but normalized as $\int d^3 p \tilde f(p) =1$. For a MB distribution, Eq.~(\ref{cr}) can be simplified as~\cite{Zhang:2017esm}
\begin{eqnarray}
\left(\frac{dN_{coll}}{dt}\right)_{MB} &=& \frac{1}{2}V\rho^2\sigma \frac{1}{4m^4 T K^2_2(m/T)} \notag\\
&\times& \int_{2m}^\infty d\sqrt{s} s (s-4m^2) K_1(\sqrt{s}/T),
\end{eqnarray}
while the collision rate $(dN_{coll}/dt)_{FD}$ for a FD distribution has to be calculated numerically through a 2-dimensional integral after integrating analytically the polar angle.

\begin{figure}[!h]
\includegraphics[width=0.8\linewidth]{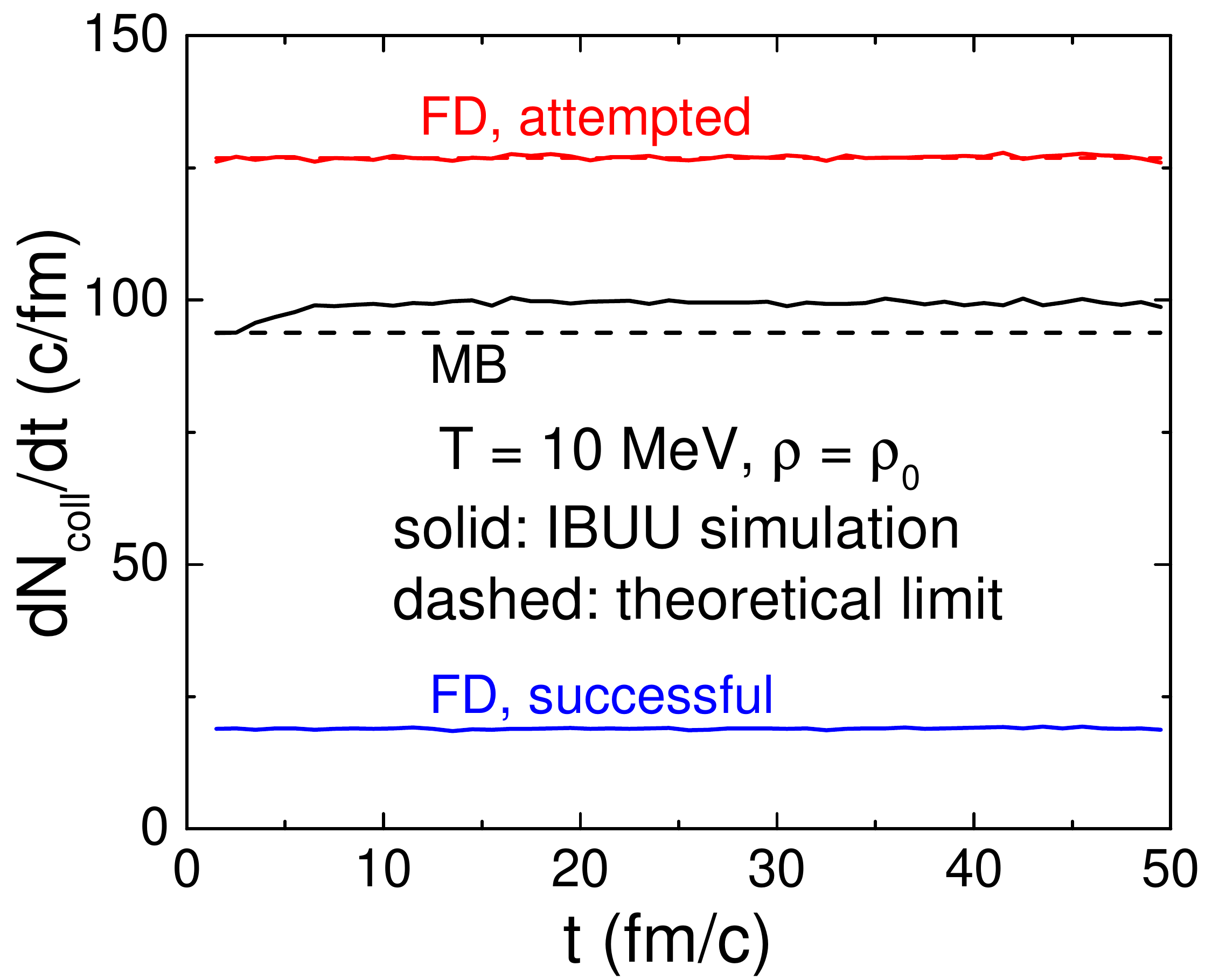}
\caption{\label{collrate} Comparison of the nucleon-nucleon collision rate from IBUU simulations in a box system with the theoretical limits. In IBUU simulations, the nucleon momenta follow a Maxwell-Boltzmann distribution or a Fermi-Dirac distribution at the density $\rho=\rho_0$ and temperature $T=10$ MeV. }
\end{figure}

Figures~\ref{collrate} compares the collision rates from IBUU simulations in the box system with the theoretical limits from Eq.~(\ref{cr}) at the density $\rho=\rho_0$ and temperature $T=10$ MeV. Without Pauli blocking, the initial MB distribution is maintained with collisions, and the collision rate after a short relaxation time is slightly higher than the theoretical limit $(dN_{coll}/dt)_{MB}=93.8$ $c$/fm, as a result of the higher-order correlations induced by collisions, consistent with the IBUU result in Ref.~\cite{Zhang:2017esm}. With Pauli blocking, the initial FD distribution is maintained with collisions, and the attempted collision rate is consistent with the theoretical limit $(dN_{coll}/dt)_{FD}=126.8$ $c$/fm, since most of attempted collisions are Pauli blocked and the higher-order correlations are not as important as in the case without Pauli blocking. As we use the analytical FD expression Eq.~(\ref{eq20}) to calculate the Pauli blocking probability $1-(1-f_1)(1-f_2)$, the successful collision rate, which dominates the shear viscosity, is reliable once the attempted collision rate is correctly reproduced.

\subsection{Dynamics in the spinodal region}

In order to use the Green-Kubo method to calculate the shear viscosity in the nuclear liquid-gas mixed phase, we need to prepare a dynamically and thermally equilibrated system with nuclear clusters. As the time evolves, the density fluctuations are required to be dynamically stable, and the temperature distributions are required to be approximately uniform. To achieve this, we use the method as described in the following.

\begin{figure}[!h]
\includegraphics[width=0.8\linewidth]{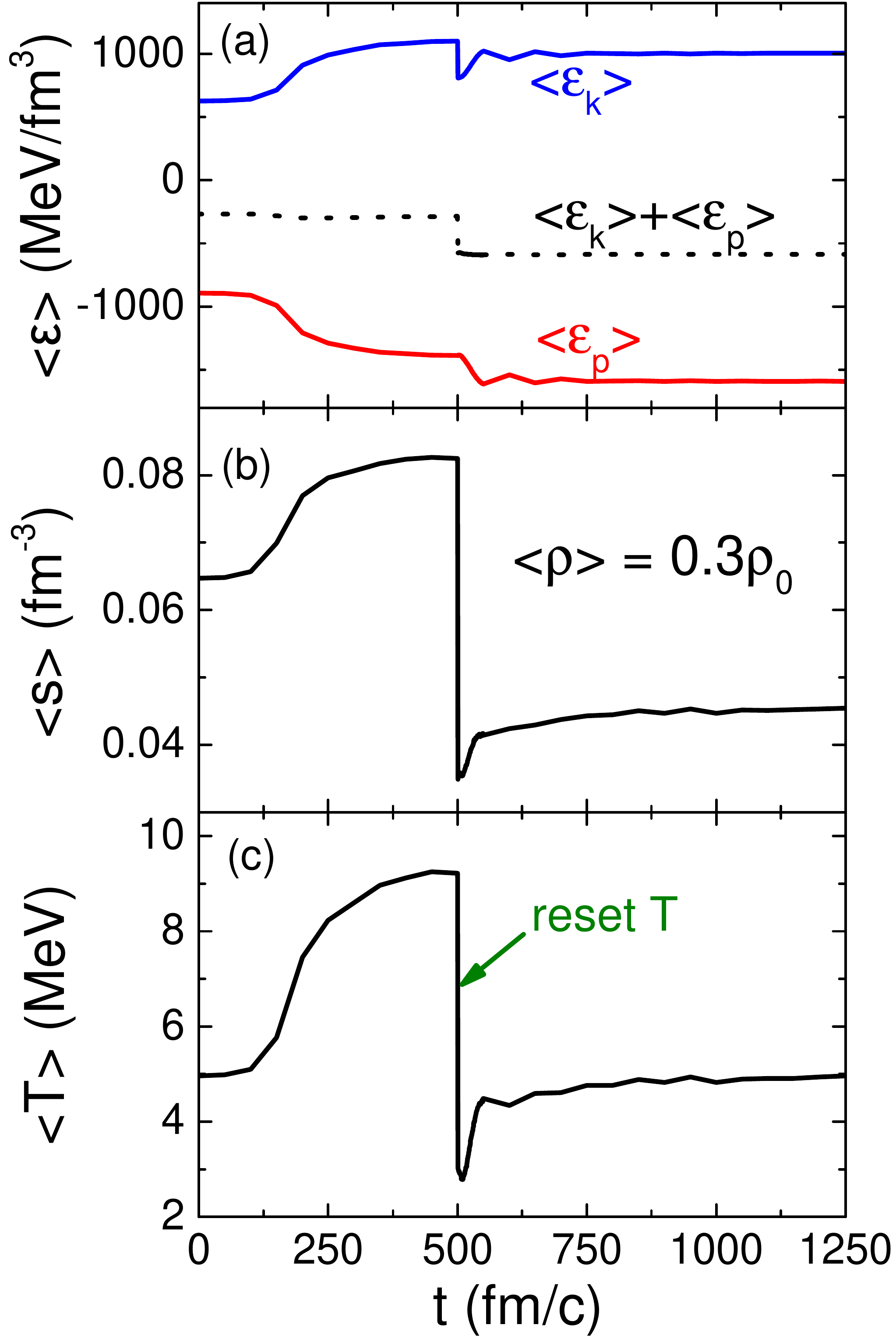}
\caption{\label{es} Time evolution of physics quantities from IBUU simulations in a box system at an average density $\langle \rho \rangle=0.3\rho_0$ and initial temperature $T=5$ MeV but with a reset of the temperature at $t=500$ fm/c. Top: Average kinetic energy density $\langle\epsilon_k\rangle$, potential energy density $\langle\epsilon_p\rangle$, and total energy density $\langle\epsilon_k\rangle+\langle\epsilon_p\rangle$; Middle: Average entropy density $\langle s \rangle$; Bottom: Average temperature $\langle T \rangle$. }
\end{figure}

\begin{figure*}[h]
\includegraphics[width=0.25\linewidth]{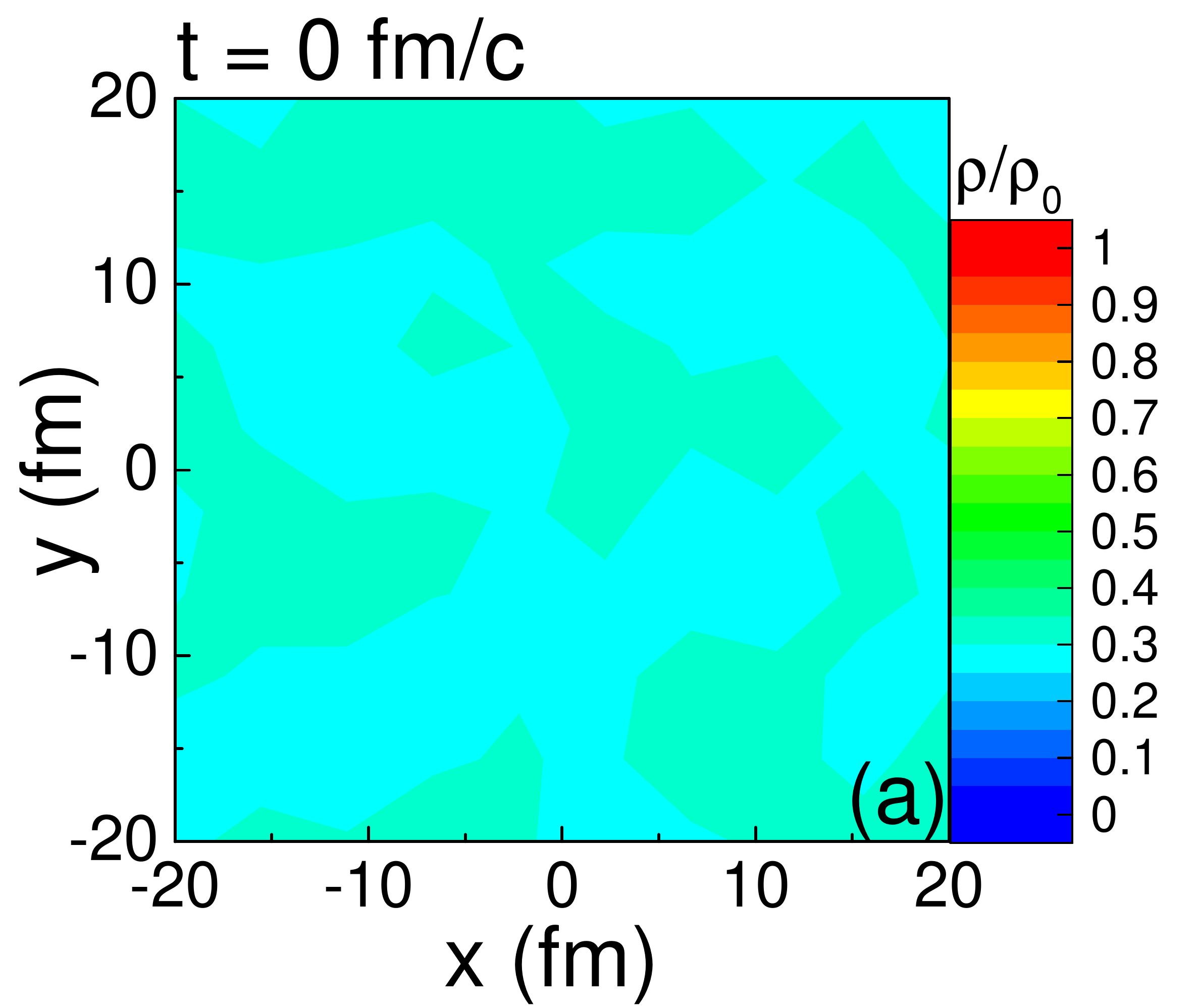}
\includegraphics[width=0.25\linewidth]{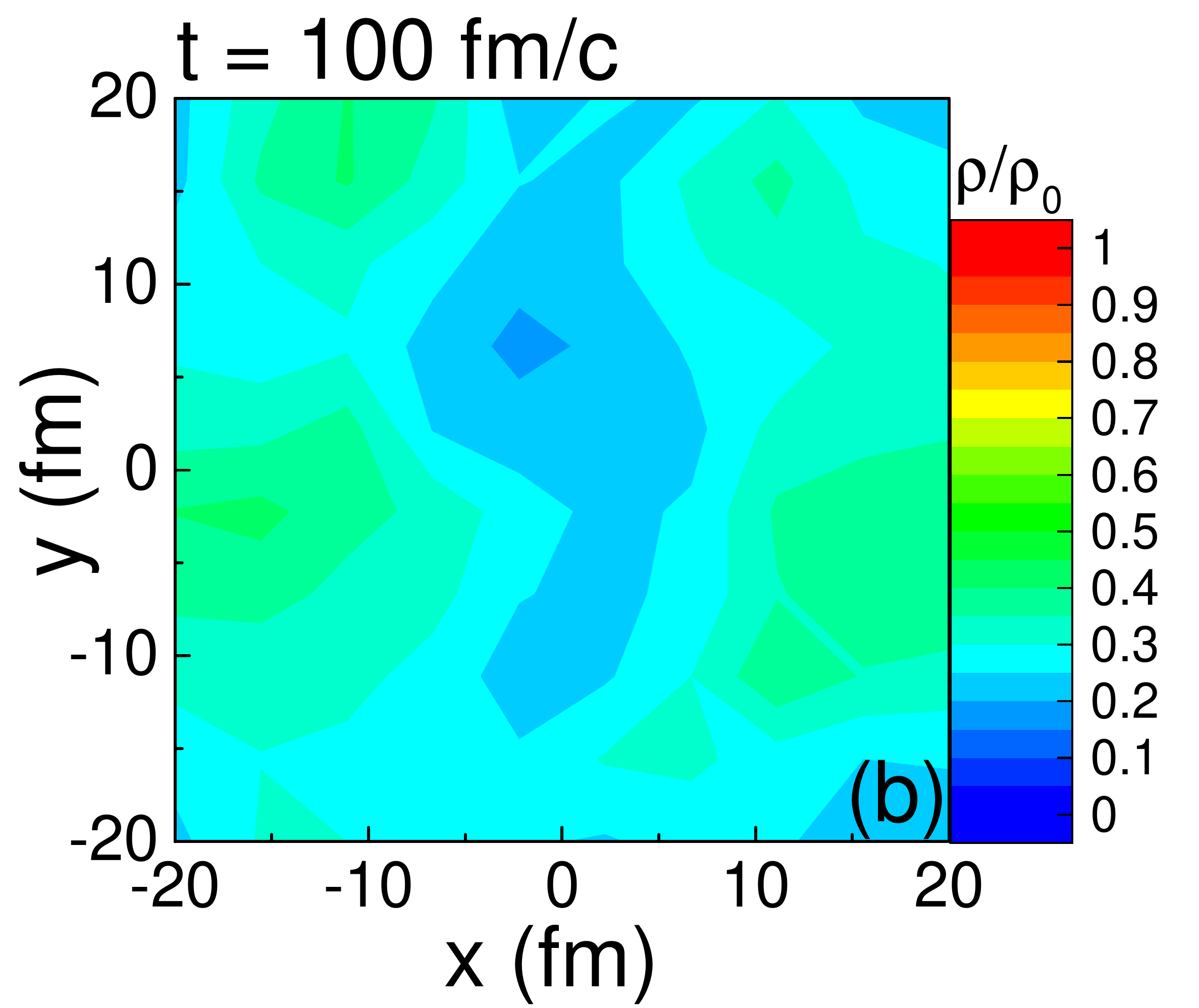}
\includegraphics[width=0.25\linewidth]{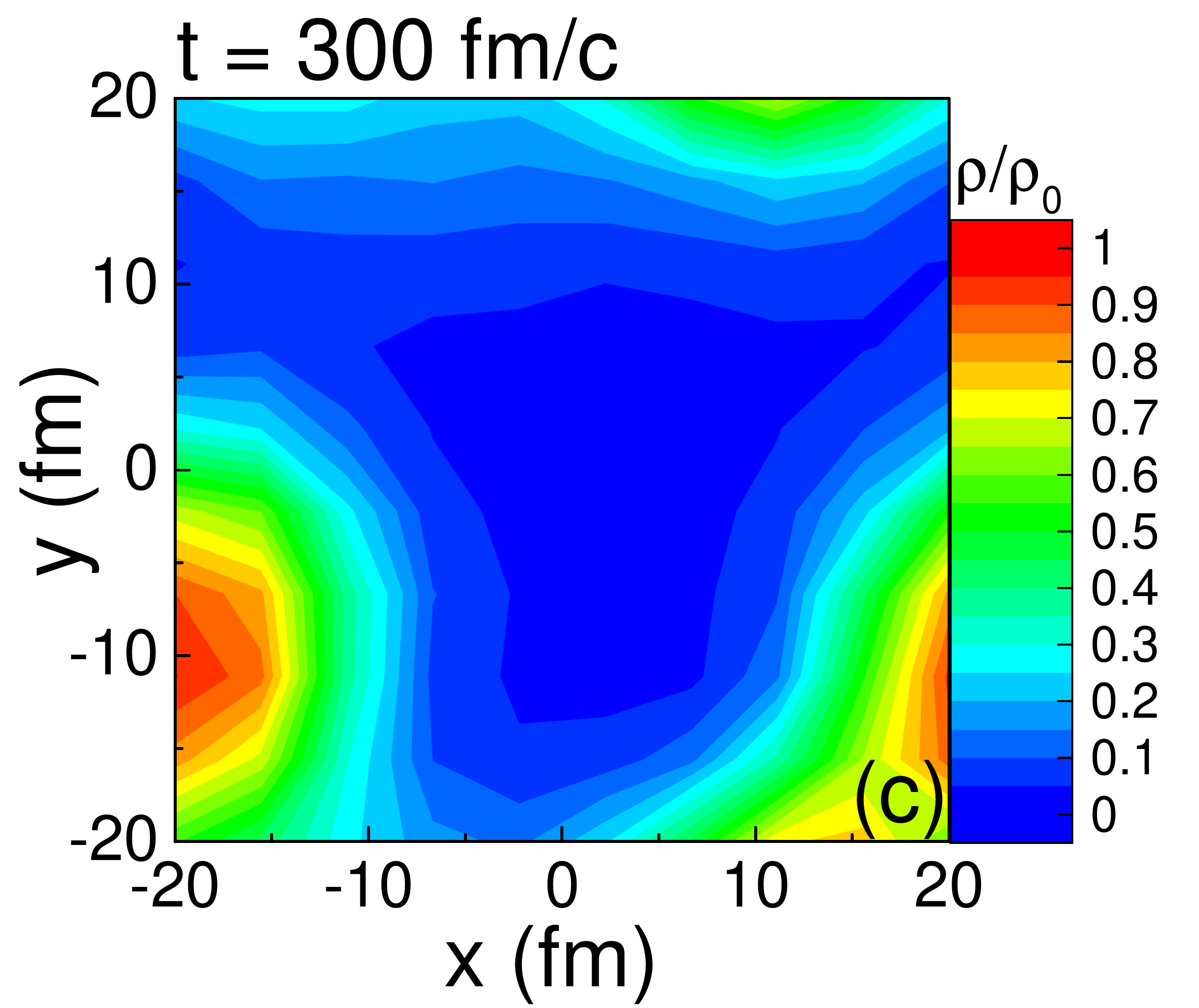}
\includegraphics[width=0.25\linewidth]{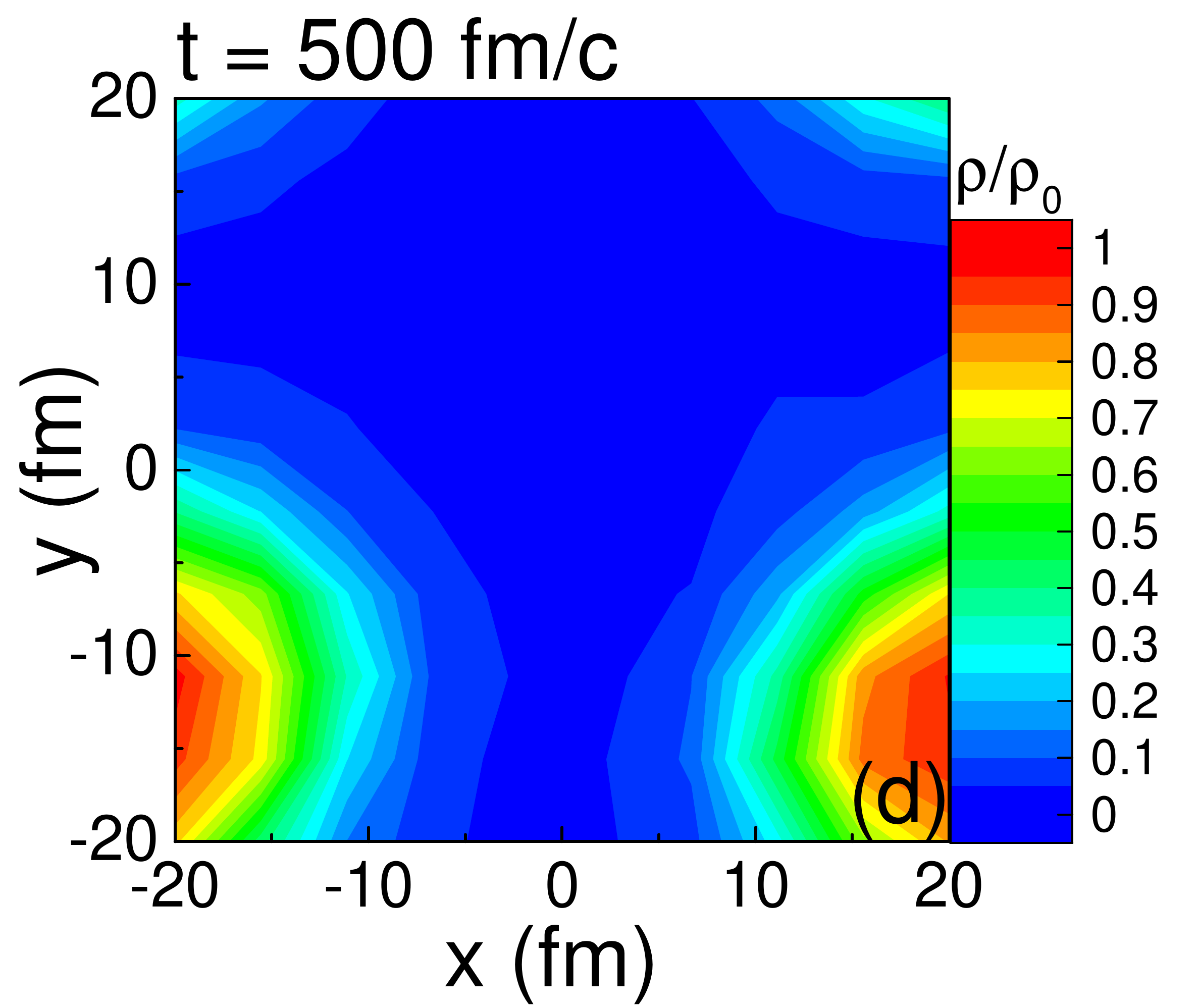}
\includegraphics[width=0.25\linewidth]{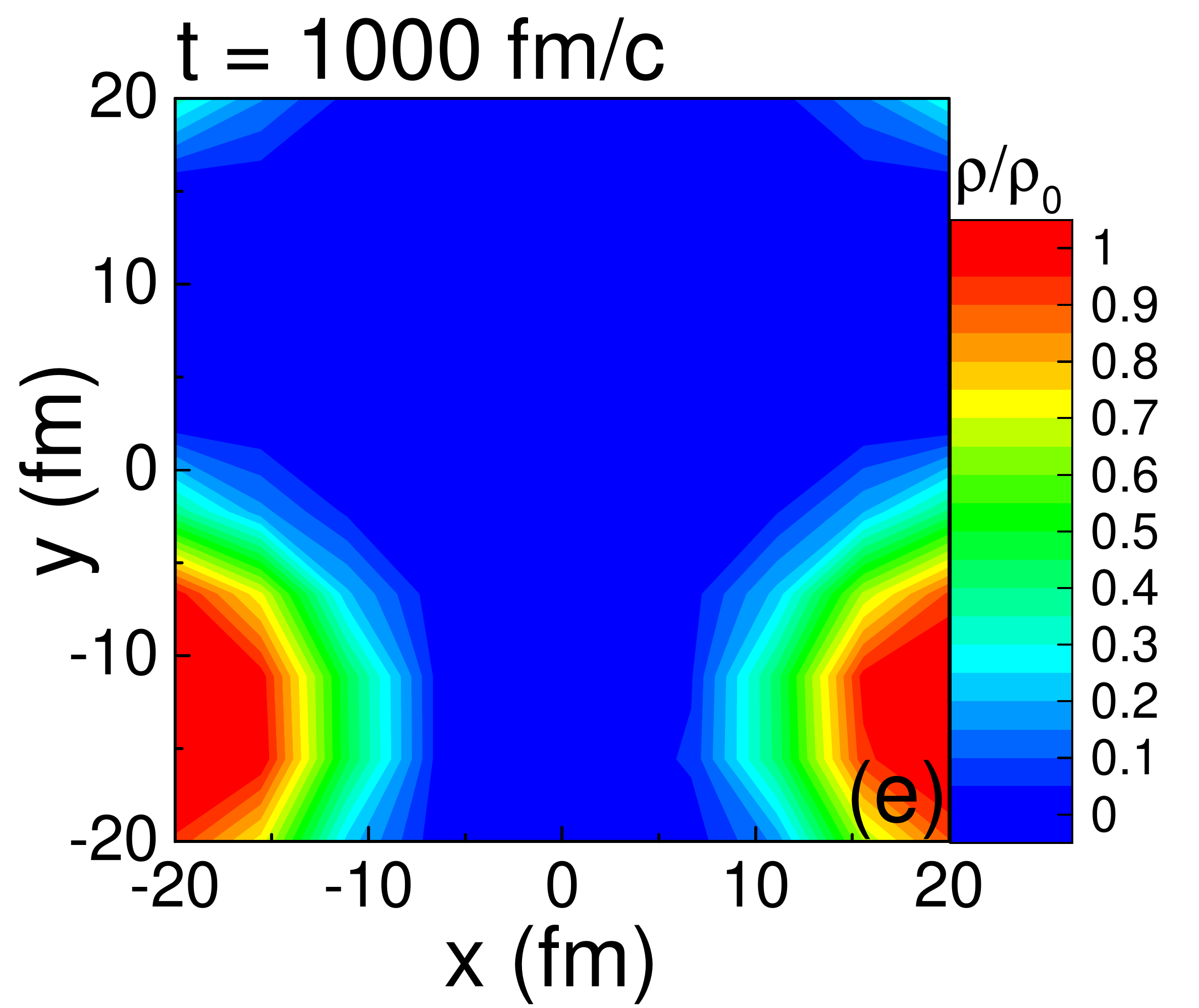}
\includegraphics[width=0.25\linewidth]{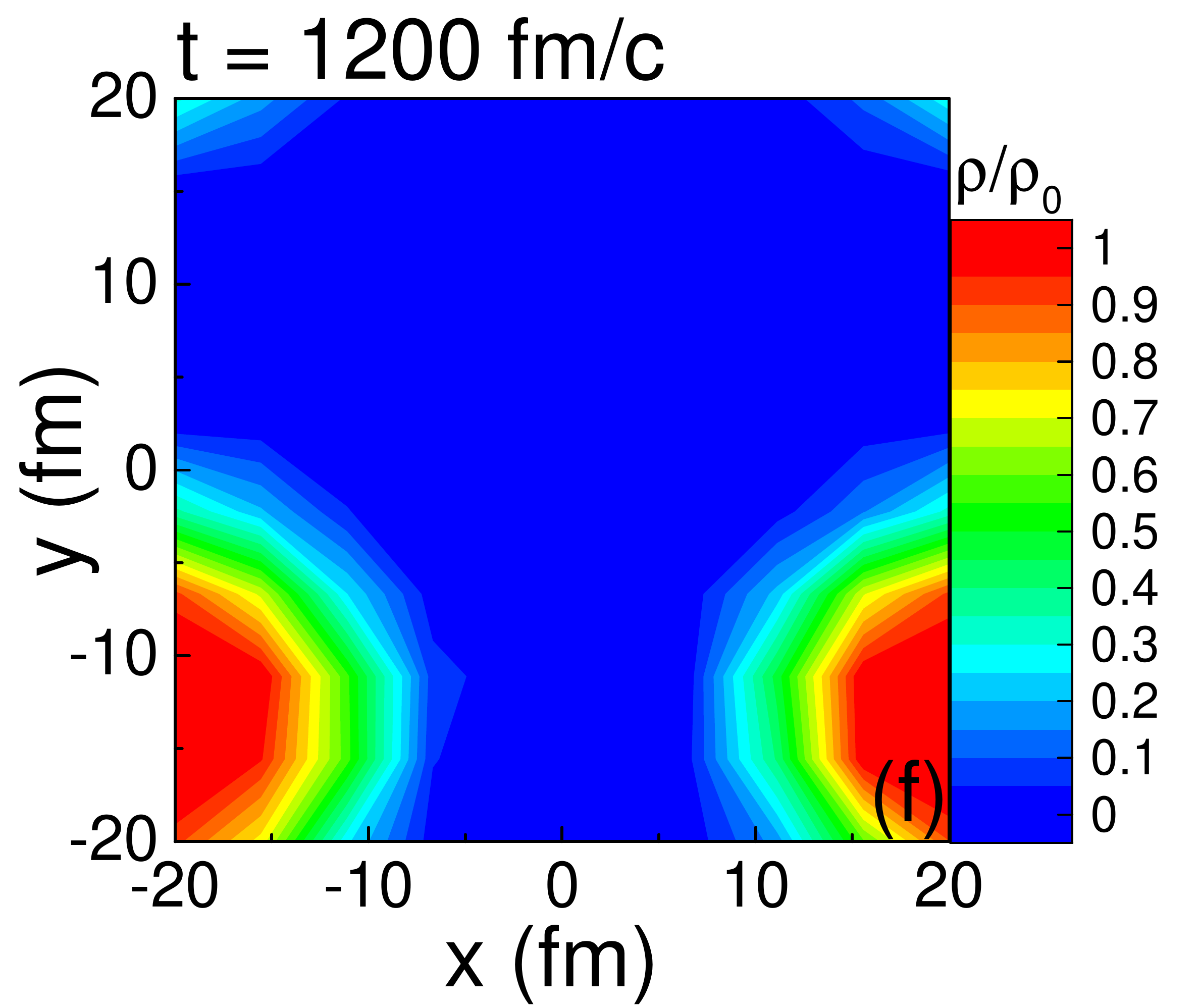}
\caption{\label{dencon} Contours of the density at different times in the x-0-y plane with $|z|<1$ fm from IBUU simulations in a box system at an average density $\langle \rho \rangle=0.3\rho_0$ and initial temperature $T=5$ MeV but with a reset of the temperature at $t=500$ fm/c.}
\end{figure*}

\begin{figure*}[h]
\includegraphics[width=0.25\linewidth]{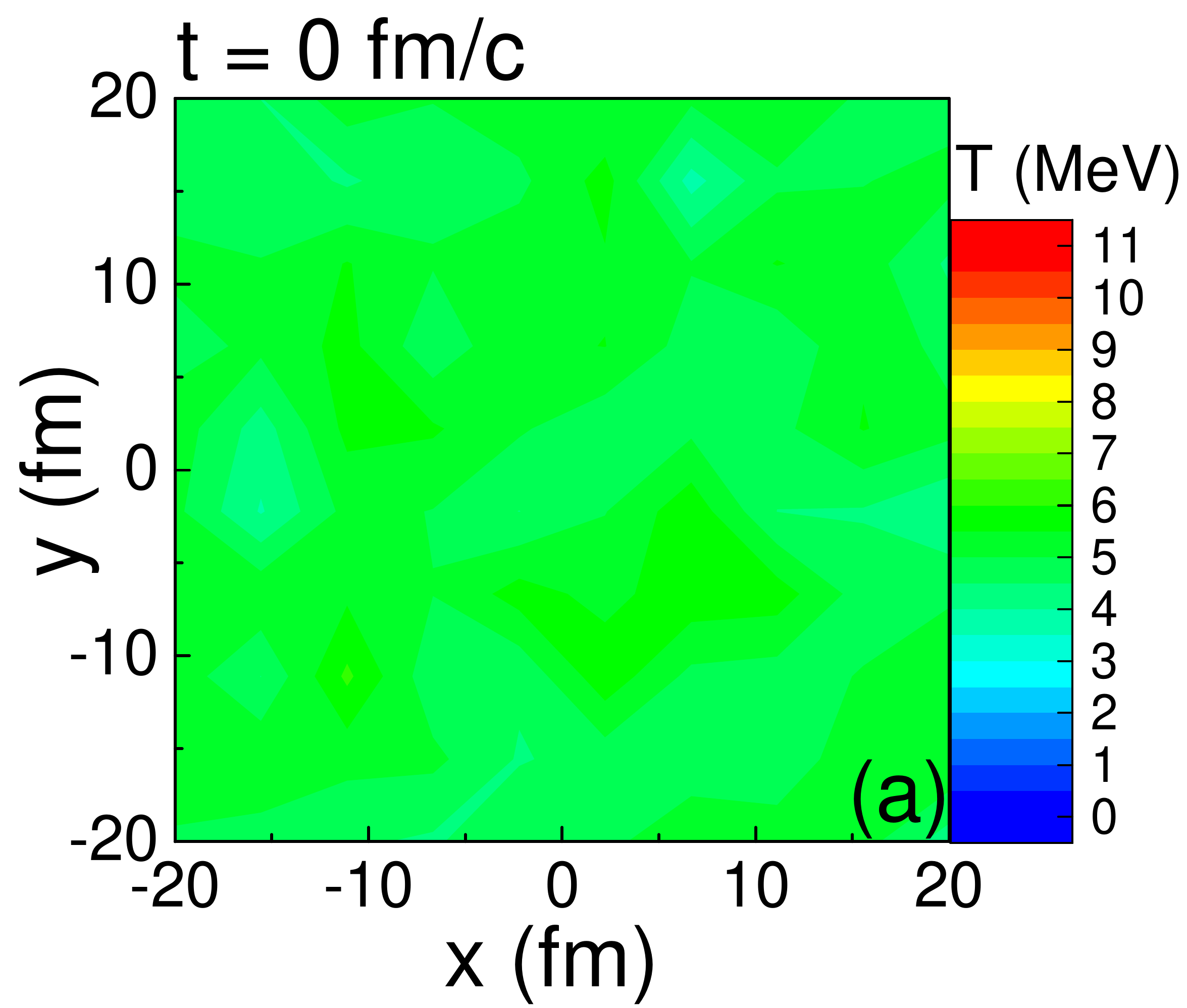}
\includegraphics[width=0.25\linewidth]{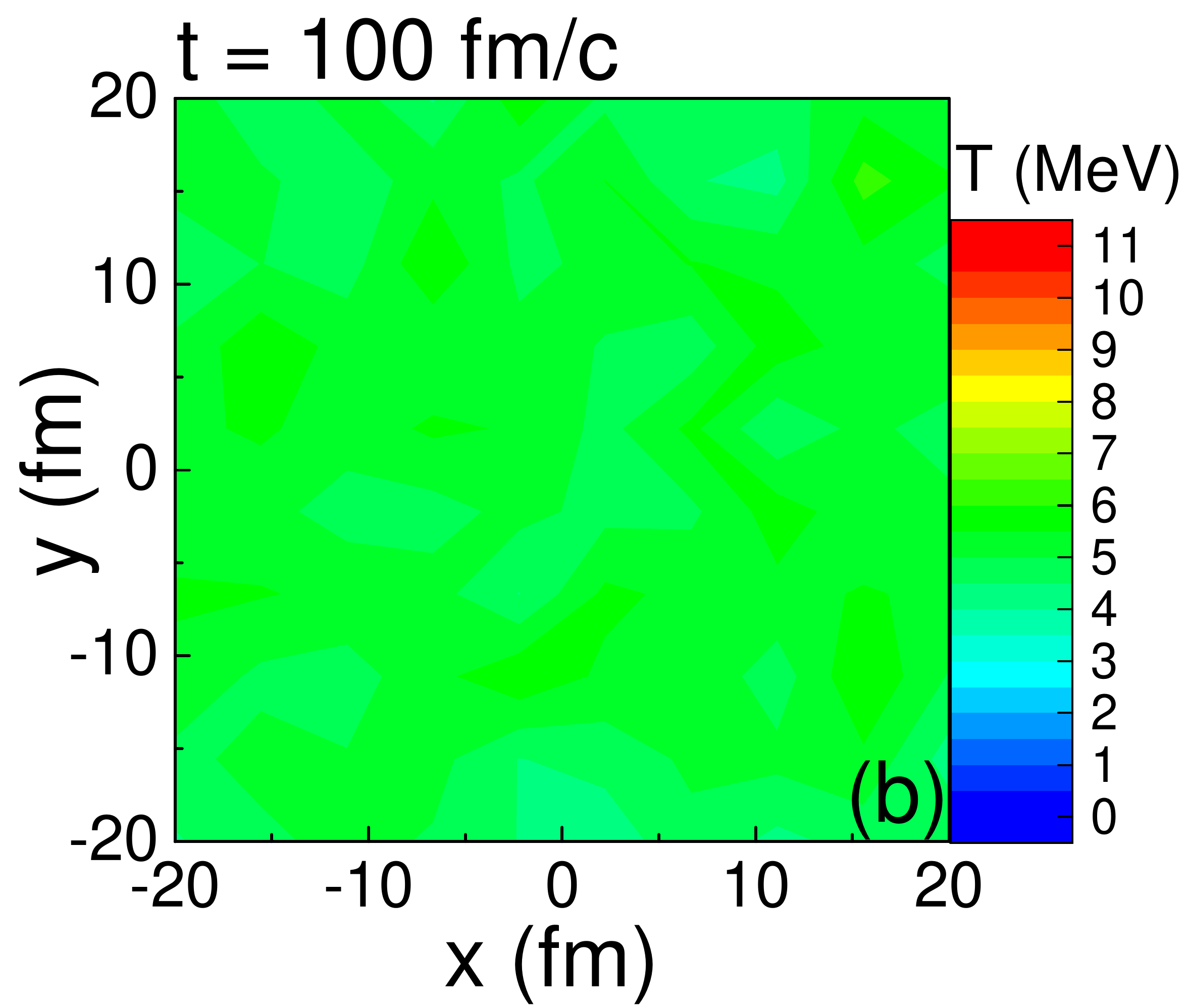}
\includegraphics[width=0.25\linewidth]{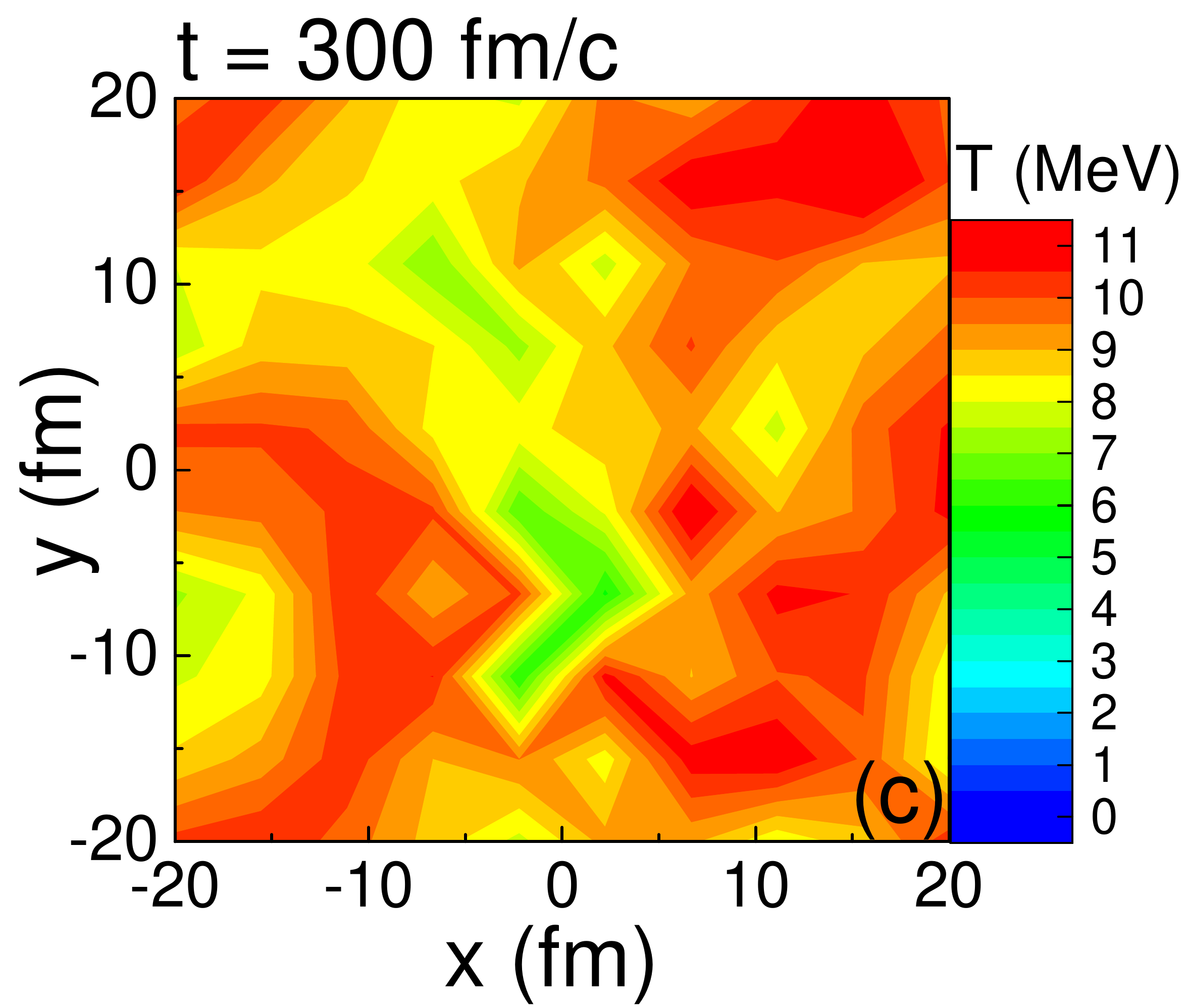}
\includegraphics[width=0.25\linewidth]{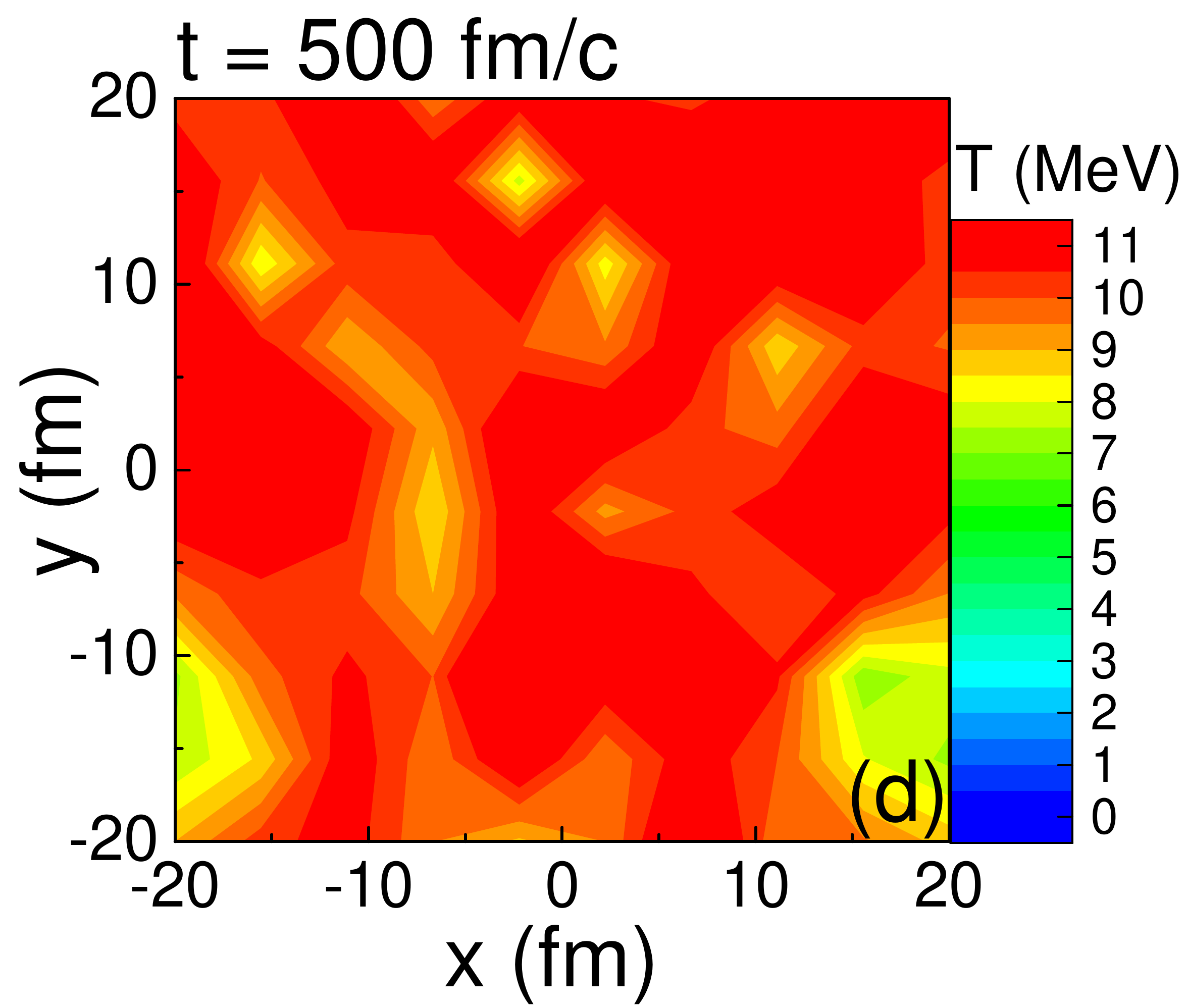}
\includegraphics[width=0.25\linewidth]{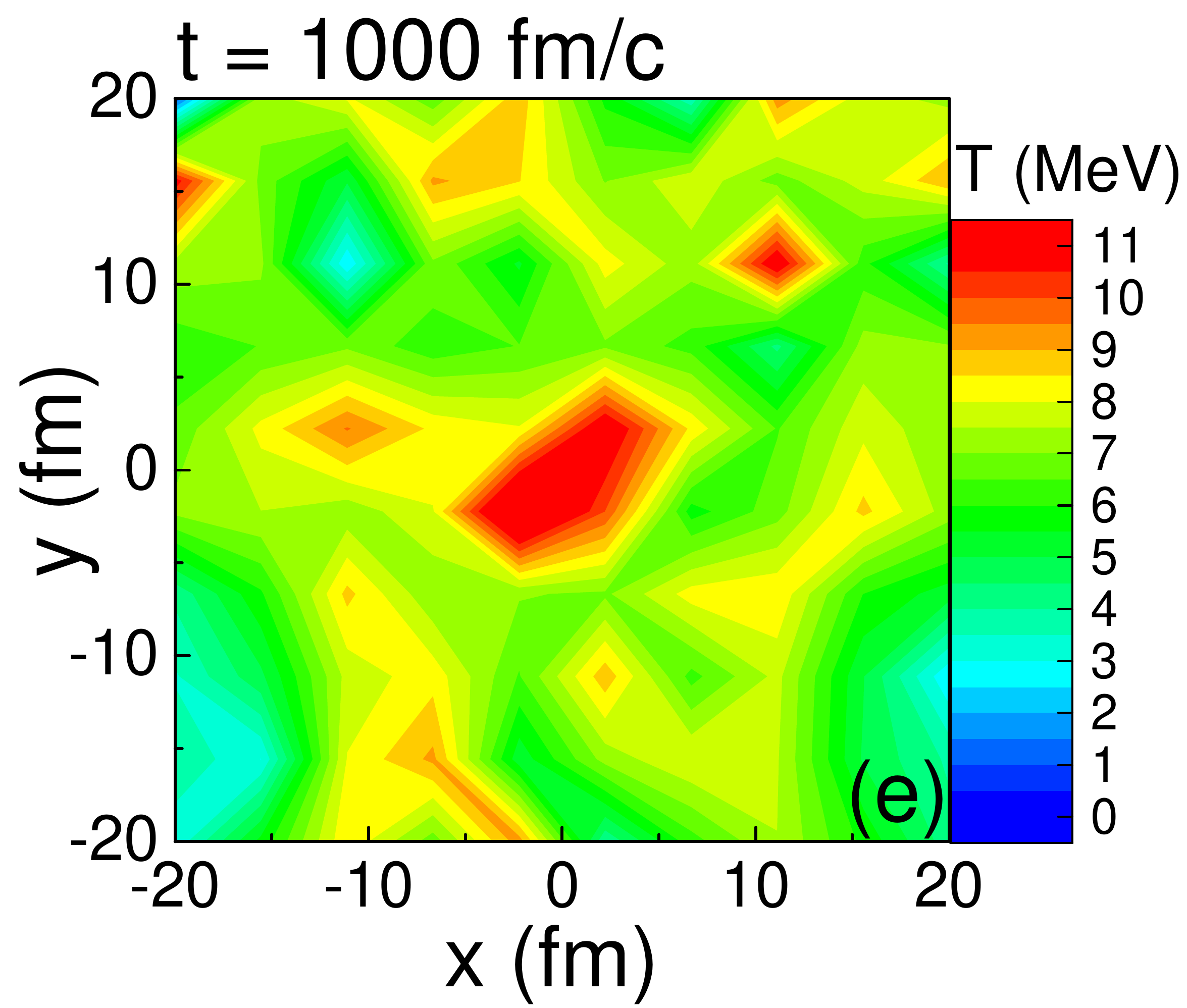}
\includegraphics[width=0.25\linewidth]{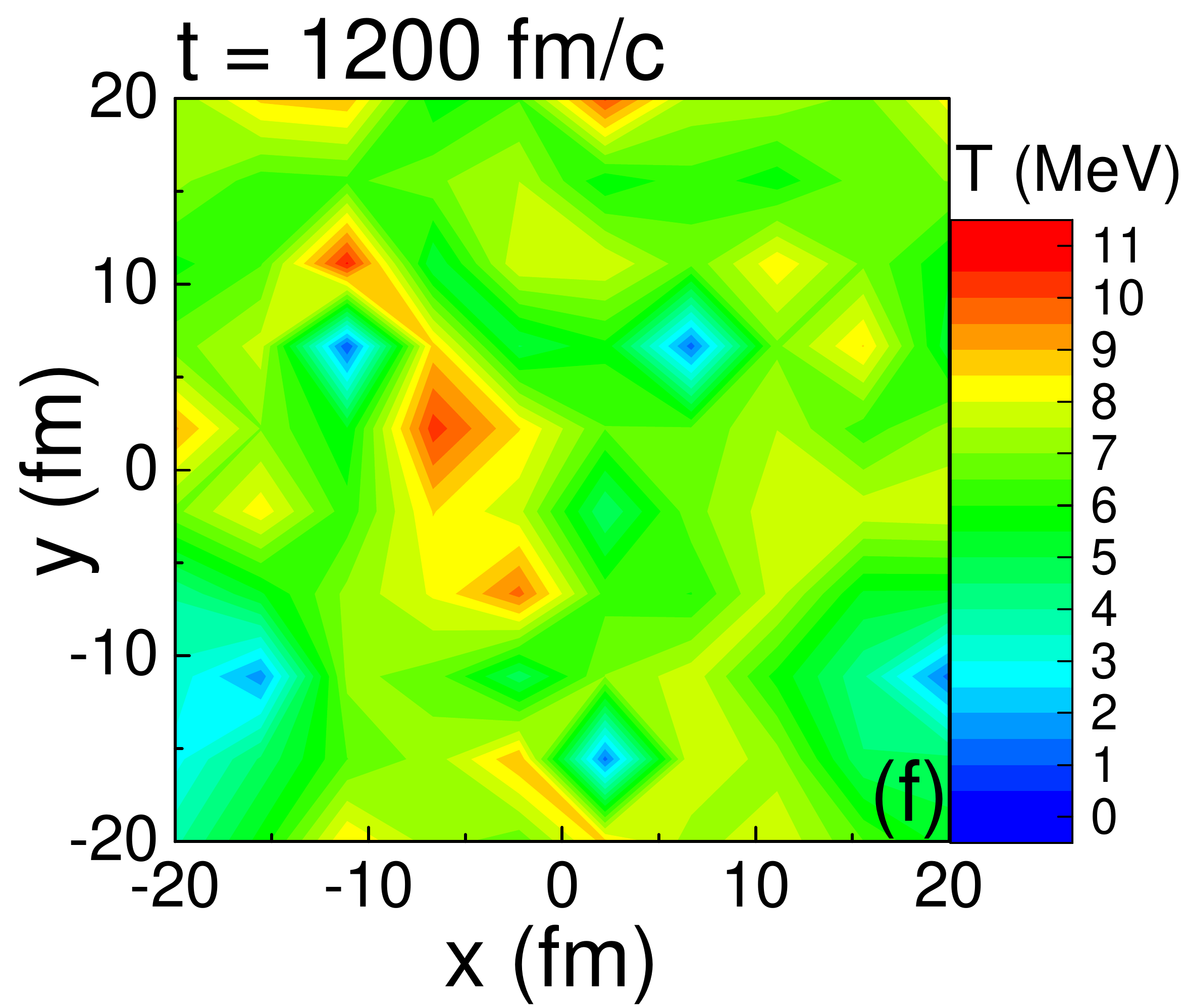}
\caption{\label{tcon} Contours of the temperature at different times in the x-0-y plane with $|z|<1$ fm from IBUU simulations in a box system at an average density $\langle \rho \rangle=0.3\rho_0$ and initial temperature $T=5$ MeV but with a reset of the temperature at $t=500$ fm/c.}
\end{figure*}

As an example, we start from a uniform nuclear matter system with a density $\rho=0.3\rho$ and an initial temperature $T=5$ MeV. The system then evolves in the box system with both the nucleon mean-field potential and nucleon-nucleon collisions, and the occupation probability for the Pauli blocking is taken to be the Fermi-Dirac distribution [Eq.~(\ref{eq20})], where the temperature $T$ and the chemical potential $\mu$ can be inversely calculated from the local kinetic energy density $\epsilon_k$ and the local number density $\rho$ according to Eqs.~(\ref{ek}) and (\ref{eq12}). In this way, the time evolution of physics quantities are displayed in Fig.~\ref{es}, and we first refer the reader to the behavior for $t<500$ fm/c. One sees that the average potential energy density $\langle \epsilon_p \rangle$ decreases with time, due to the clusterization of nucleons in the spinodal region~\cite{Burgio:1991ej}. According to the energy conservation as maintained in the simulation based on the well-established lattice Hamiltonian framework, the average kinetic energy density $\langle \epsilon_k \rangle$ increases with time, and the average temperature $\langle T \rangle$ and the average entropy density $\langle s \rangle$ also increase with time, with the $\langle T \rangle$ calculated by averaging the local temperature $T(\vec{r})$ weighted by the local density, i.e.,
\begin{equation}
\langle T \rangle = \frac{\int d^3 r \rho(\vec{r}) T(\vec{r})}{\int d^3 r \rho(\vec{r})},
\end{equation}
and the $\langle s \rangle$ calculated similarly with the local entropy energy obtained according to Eq.~(\ref{s}). The corresponding contours of the number density and the temperature at typical times in the x-0-y plane are displayed in Figs.~\ref{dencon} and \ref{tcon}, respectively. Initially, both the number density and the temperature are uniformly sampled, but with small statistical fluctuations. These fluctuations grow with time and serve as seeds for clusterization, since the $(\rho,T)$ state of the nuclear system is in the mechanical instability region. At $t=300$ fm/c, clusters are obviously formed, and the overall temperature is increased. At $t=500$ fm/c, clusterization becomes stable, and the temperature is further increased. One sees that the temperature is slightly lower at high densities compared to that at low densities.

In the above process, we start from the system at a lower temperature, but ends up with a system at a higher temperature, and with a not very satisfactory temperature distribution. By changing the initial temperature of a uniform system, we find that we are unable to achieve a system with clusters at low temperatures, e.g., $T \leq 5$ MeV. To achieve a system with dynamically stable clusterizations and a more uniform temperature distribution at $T=5$ MeV, we reset the temperature at $t=500$ fm/c and use it as a new initial state. To do this, we resample the momentum distribution of nucleons in each cell according to the Fermi-Dirac distribution [Eq.~(\ref{eq20})], with the temperature $T$ reset to be about 3 MeV in this case, and the chemical potential $\mu$ determined by the local number density. The average potential energy density $\langle \epsilon_p \rangle$ determined by the density distribution is unchanged after the reset of the temperature, while there is a sudden decrease of $\langle \epsilon_k \rangle$, $\langle \epsilon_k \rangle+\langle \epsilon_p \rangle$, and $\langle s \rangle$, as shown in Fig.~\ref{es}. To achieve a more uniform temperature distribution in the subsequent box simulations, we use the average temperature to calculate the occupation probability in the Pauli blocking factor, where the chemical potential is determined by the local density. Since the clusterization effect is even stronger at a lower reset temperature, $\langle \epsilon_p \rangle$ decreases but $\langle \epsilon_k \rangle$, $\langle s \rangle$, and $\langle T \rangle$ increase during a short time after the reset of the temperature. Afterwards, the system gradually evolves to a dynamically stable state, since all physics quantities remain almost unchanged in the later process, as seen from Fig.~\ref{es}. From the corresponding contours of the number density and the temperature as shown in Figs.~\ref{dencon} and \ref{tcon}, it is seen that the clusterization is only slightly enhanced after the temperature is reset, and this leads to only a weak correlation between the density distribution and the temperature distribution. By resetting the temperature for additional times and with more test particles, a liquid-gas mixed system with a more uniform temperature distribution can be obtained, while we expect that the results of the shear viscosity remain almost unchanged.

\begin{figure}[!h]
\includegraphics[width=0.8\linewidth]{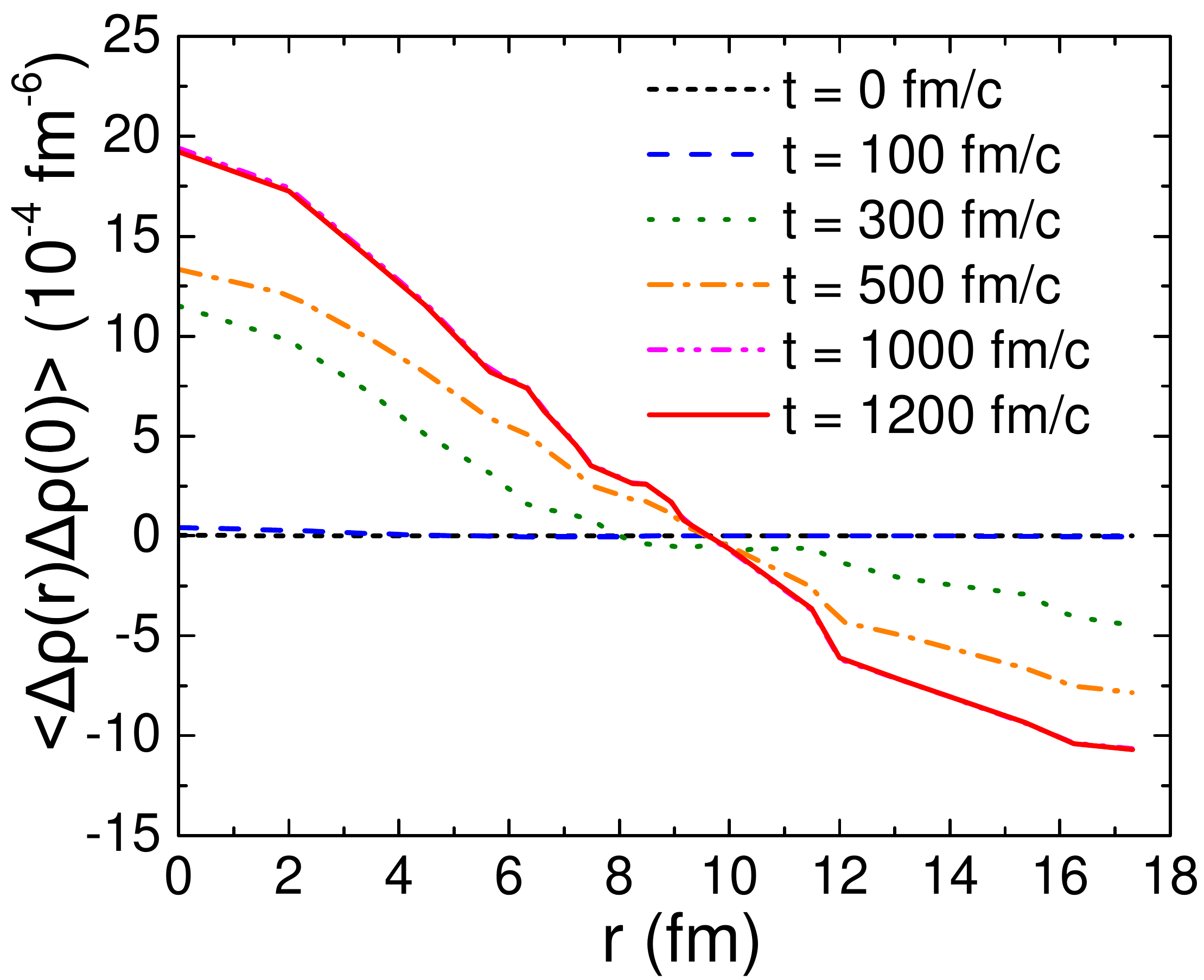}
\caption{\label{corr} Correlations of density fluctuations at different times from IBUU simulations in a box system corresponding to the same evolution in Figs.~\ref{dencon} and \ref{tcon}.}
\end{figure}

We also display in Fig.~\ref{corr} the correlations $\langle \Delta \rho(r) \Delta \rho(0) \rangle$ of density fluctuations at different times in the same system as Figs.~\ref{dencon} and \ref{tcon}, where $\Delta \rho(r)=\rho(r)-\langle \rho \rangle$ represents the density fluctuation in a cell with a distance $r$ from the original cell, with $\langle \rho \rangle$ being the average density of the box system. The periodic boundary condition is taken into account in evaluating the correlation of density fluctuations, so the maximum value of $r$ is $\sqrt{3(L/2)^2} \approx 17.3$ fm. Initially, there is no density fluctuation and thus zero correlation. As time evolves, with the appearance of clusterization, the correlation of the density fluctuation grows at $t=300$ and 500 fm/c. After the temperature is reset at $t=500$ fm/c, there are small modifications on the density fluctuation, and the correlation further grows and becomes saturated at $t=1000$ and 1200 fm/c. For the density fluctuations at later times as shown in Fig.~\ref{dencon}, the correlation of density fluctuations is positive for small $r$ corresponding to the liquid drop of nucleons with a centain volume, and negative for large $r$ corresponding to the gas phase away from the liquid drop. The radius of the cluster can be estimated as the half-height width of $\langle \Delta \rho(r) \Delta \rho(0) \rangle$, which is about 5 fm, consistent with the observation from Fig.~\ref{dencon}.

From monitoring the time evolutions of the density distribution, the average kinetic and potential energy density, the average entropy density, the average temperature, and the correlation of density fluctuations, we found that the dynamical equilibrium is completely reached after $t=1200$ fm/c. We thus set $t_0=1200$ fm/c as the starting time for the calculation of the shear viscosity based on the Green-Kubo method. For other ($\rho$, $T$) states in the spinodal region as in Fig.~\ref{prhoT} (b), a similar process is used to achieve dynamic and thermal equilibrium, while $t_0$ can be slightly different. Since in non-uniform systems we mostly talk about average quantities, the average symbol ``$\langle ... \rangle$'' will be omitted in most cases of the subsequent discussions.

\subsection{Specific shear viscosity}

\begin{figure}[!h]
\includegraphics[width=0.8\linewidth]{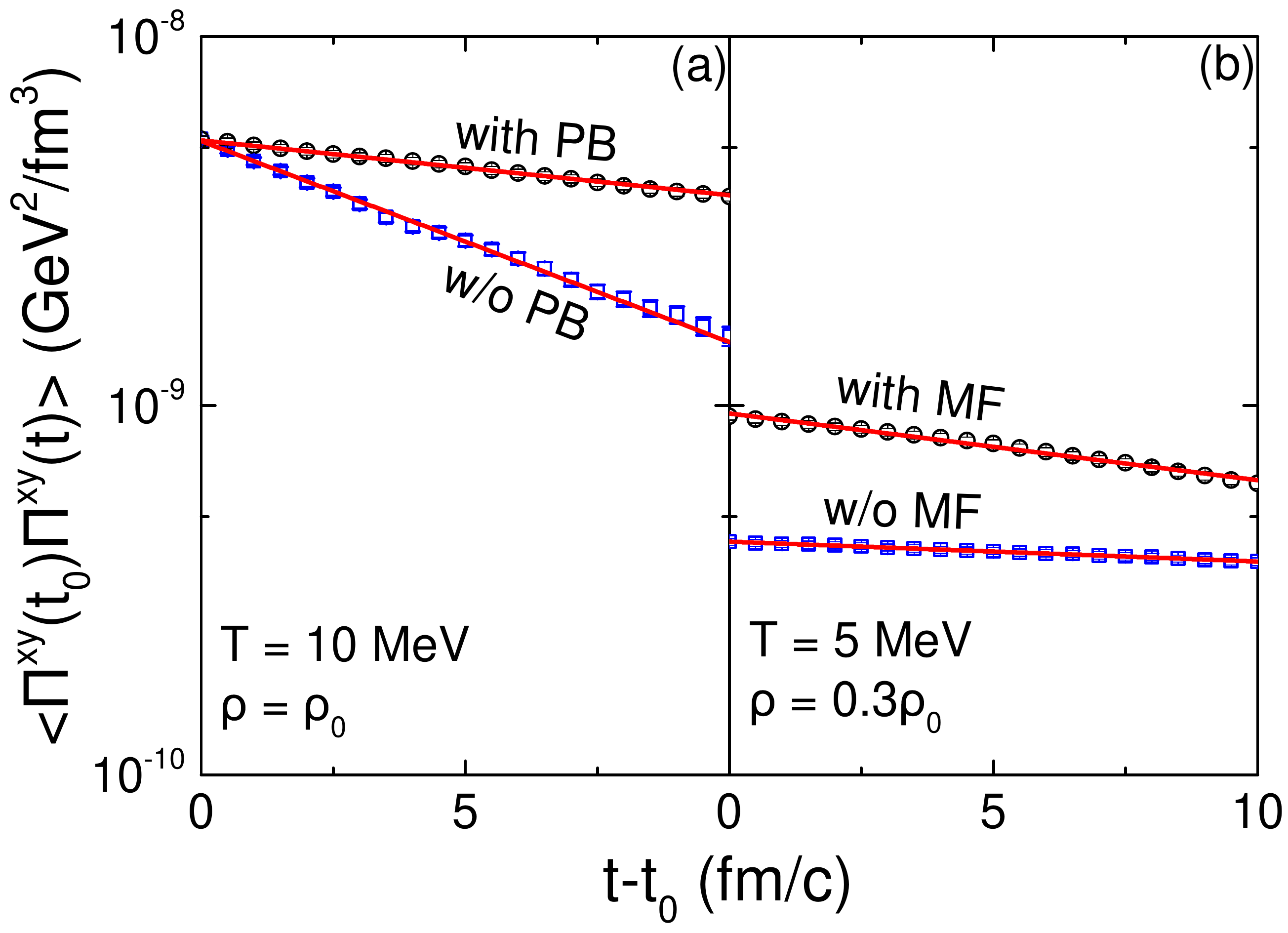}
\caption{\label{pai} Correlations of the energy-momentum tensor as a function of time from IBUU simulations in a box system. Left: Uniform system at a density $\rho=\rho_0$ and temperature $T=10$ MeV; Right: Uniform (w/o MF) and non-uniform (with MF) system at an average density $\rho=0.3\rho_0$ and temperature $T=5$ MeV.}
\end{figure}

The shear viscosity is calculated based on the Green-Kubo method by evaluating the correlation of the energy-momentum tensor according to Eq.~(\ref{eq2s}), whose time evolutions for typical systems from IBUU simulations are displayed in Fig.~\ref{pai} for illustration, based on the statistical average of about 10000 events for each case. Figure~\ref{pai} (a) displays the results from a uniform system at a density $\rho=\rho_0$ and temperature $T=10$ MeV out of the spinodal region, and IBUU simulations are performed with and without Pauli blocking (PB). The correlations of the energy-momentum tensor in the two cases start from the same value, and then decrease exponentially with time. The decreasing trend reflects how fast the system forgets its initial state, and it is stronger for the case without PB due to more successful nucleon-nucleon collisions compared to the case with PB. Figure~\ref{pai} (b) compares the results at an average density $\rho=0.3\rho_0$ and temperature $T=5$ MeV with and without the mean-field potential (MF), corresponding to non-uniform and uniform systems, respectively. The correlation of the energy-momentum tensor for a non-uniform system starts from a larger value due to the enhanced correlation from the clusterization, and decreases exponentially with time more rapidly as a result of more successful nucleon-nucleon collisions within high-density clusters, compared to the case for a uniform system. Using the least square fit method, the function $\langle\Pi^{x y}(t_0) \Pi^{x y}(t)\rangle$ can be parameterized as
\begin{equation}
\label{eq5}
\langle\Pi^{x y}(t_0) \Pi^{x y}(t)\rangle  = Ae^{-B(t-t_0)},
\end{equation}
where $A$ is determined by the correlation of the energy-momentum tensor at $t=t_0$, and $B$ reflects how rapidly the correlation decreases. According to Eq.~(\ref{eq2s}), the shear viscosity can then be expressed as
\begin{equation} \label{etaab}
\eta = \frac{AV}{BT},
\end{equation}
where a larger $B$ from a stronger collision effect reduces the value of $\eta$.

\begin{figure}[!h]
\includegraphics[width=1\linewidth]{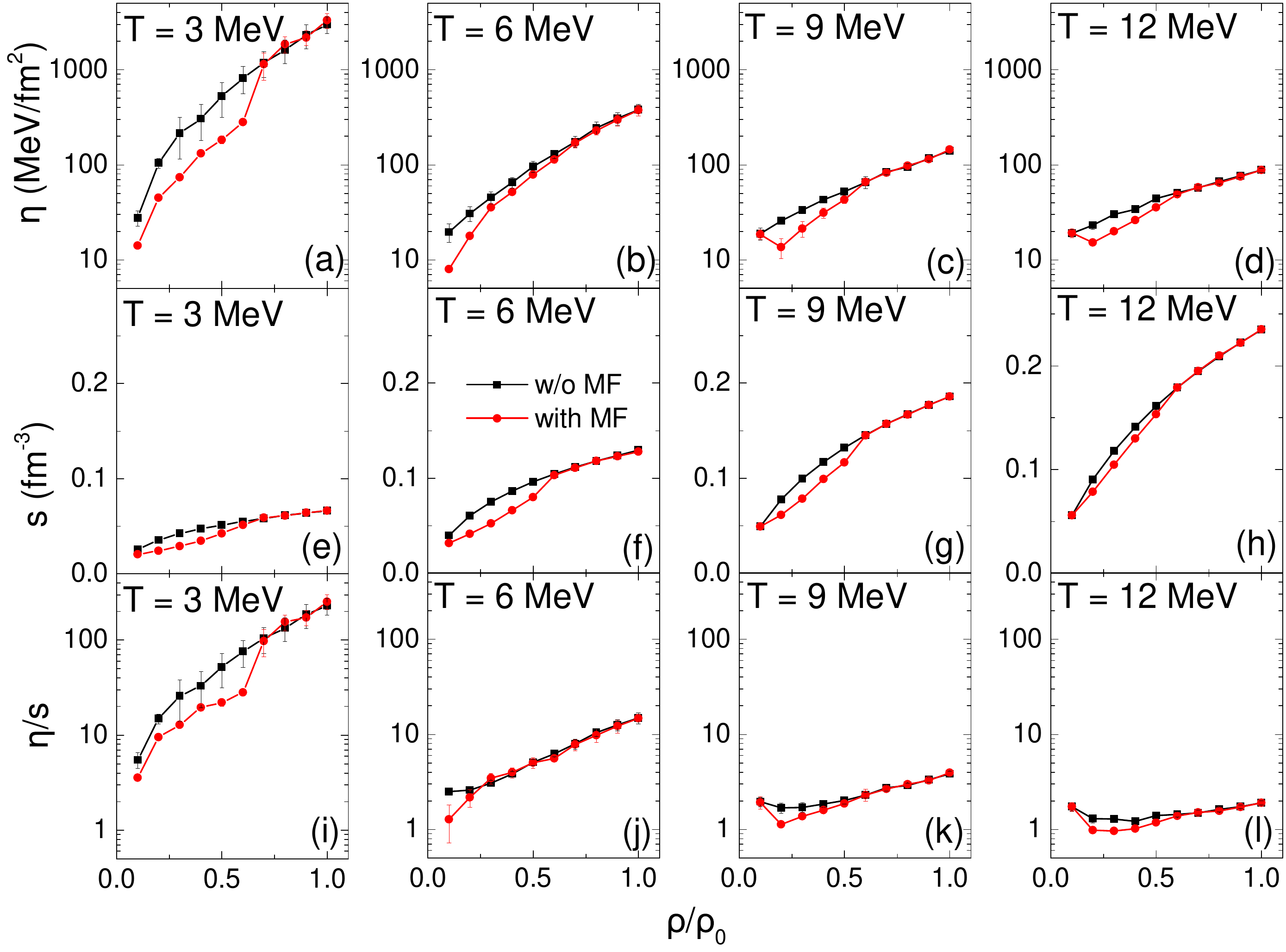}
\caption{\label{etasT} Shear viscosity $\eta$ (first row), entropy density $s$ (second row), and specific shear viscosity $\eta/s$ (third row) as a function of average density $\rho$ from IBUU simulations in a box system at the temperatures $T=3$ (first column), 6 (second column), 9 (third column), and 12 MeV (fourth column) with and without mean-field potential (MF). }
\end{figure}

\begin{figure}[!h]
\includegraphics[width=1\linewidth]{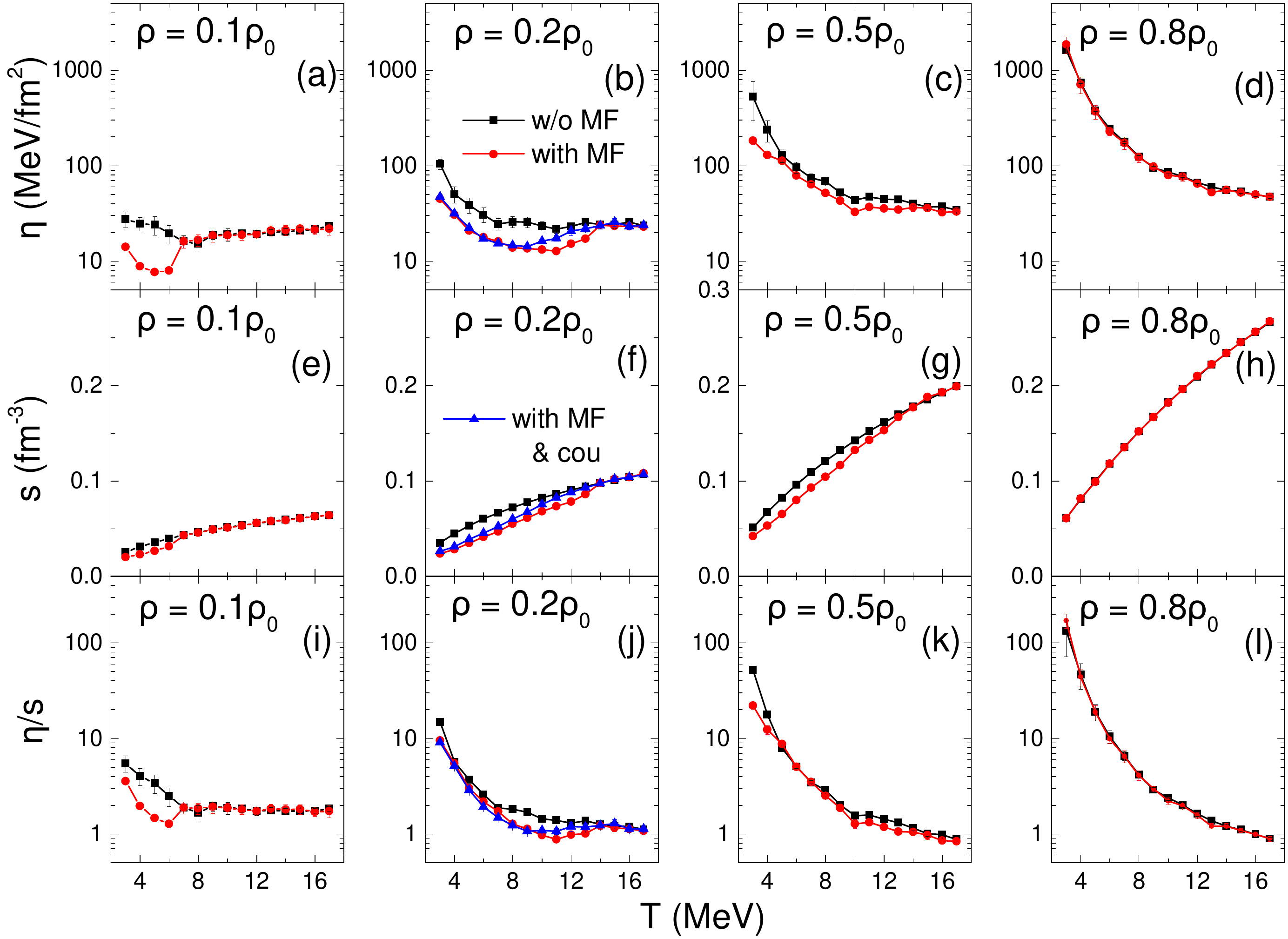}
\caption{\label{etasrho} Shear viscosity $\eta$ (first row), entropy density $s$ (second row), and specific shear viscosity $\eta/s$ (third row) as a function of temperature $T$ from IBUU simulations in a box system at average densities $\rho=0.1\rho_0$ (first column), $0.2\rho_0$ (second column), $0.5\rho_0$ (third column), and $0.8\rho_0$ (fourth column) with and without mean-field potential (MF). Results with both the mean-field potential and the Coulomb force (MF \& cou) are compared for the average density $\rho=0.2\rho_0$.}
\end{figure}

Figures~\ref{etasT} and \ref{etasrho} display, respectively, the shear viscosity $\eta$, the average entropy density $s$, and the specific shear viscosity $\eta/s$ as a function of average density $\rho$ at different temperatures and as a function of temperature $T$ at different average densities, where results with and without mean-field potential are compared. The error bars mostly originate from the fitting error according to Eq.~(\ref{eq5}). The shear viscosity generally increases with increasing average density due to the stronger Pauli blocking effect at higher densities, and decreases with increasing temperature due to the weaker Pauli blocking effect and thus more successful collisions at higher temperatures. The average entropy density generally increases with both increasing density and temperature, as a result of a more populated phase space at higher densities or temperatures. The ratio of the shear viscosity to the average entropy density, i.e., the specific shear viscosity, mostly decreases with increasing temperature for a given average density, but increases with the increasing average density for a given temperature. For the average density and temperature out of the spinodal region as shown in Fig.~\ref{prhoT} (b), the results with mean-field potential agree with those without mean-field potential within error bars, since the system is always uniform. For the average density and temperature inside the spinodal region, the system is non-uniform (uniform) with (without) mean-field potential. The formation of high-density hot clusters in the spinodal region enhances the collision effect and thus reduces the shear viscosity, while the average entropy density is reduced in non-uniform systems compared to that in uniform systems. Taking the ratio of $\eta$ to $s$, the specific shear viscosity is seen to be reduced in non-uniform systems compared to uniform systems at the same average densities and temperatures.

The minimum of the specific shear viscosity as a function of temperature is of special interest, and in the present framework it is seen only at very low average densities. At an average density $\rho=0.1\rho_0$, a minimum $\eta/s$ is seen at about $T=6$ MeV. At an average density $\rho=0.2\rho_0$, a minimum $\eta/s$ is seen at about $T=11$ MeV. At even higher average densities, the minimum of $\eta/s$ is not obviously seen. At a given temperature, the density dependence of $\eta/s$ may also show a minimum behavior at higher temperatures, and they are around $\rho=0.2\rho_0$ at both $T=9$ MeV and 12 MeV.

Once the Coulomb force is incorporated, the clustering effect becomes weaker compared to what has been shown in Sec.~\ref{sec:results} B. This is understandable, since the energy conservation condition leads to a reduced maximum density of clusters due to the repulsive nature of the Coulomb potential for protons. The effect of incorporating the Coulomb force on the shear viscosity is illustrated in Fig.~\ref{etasrho} for the average density of $\rho=0.2\rho_0$. It is seen that the reduced clustering effect in the presence of the Coulomb force increases both $\eta$ and $s$ at about $T=9 \sim 13$ MeV, leading to a slightly increased $\eta/s$ and a lower temperature for the minimum $\eta/s$, while the qualitative behaviors of these quantities remain generally unchanged.

\begin{figure}[!h]
\includegraphics[width=0.8\linewidth]{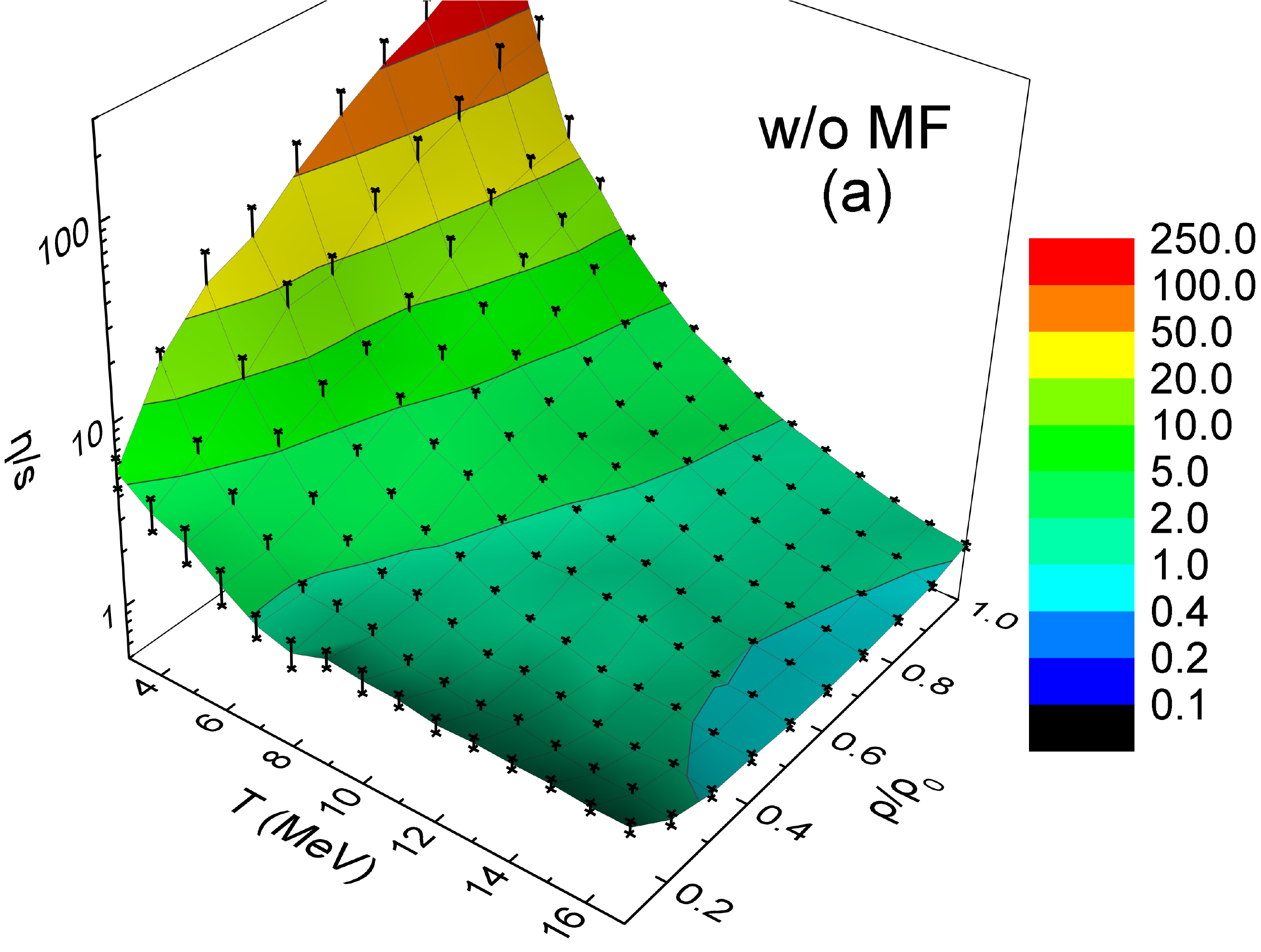}
\includegraphics[width=0.8\linewidth]{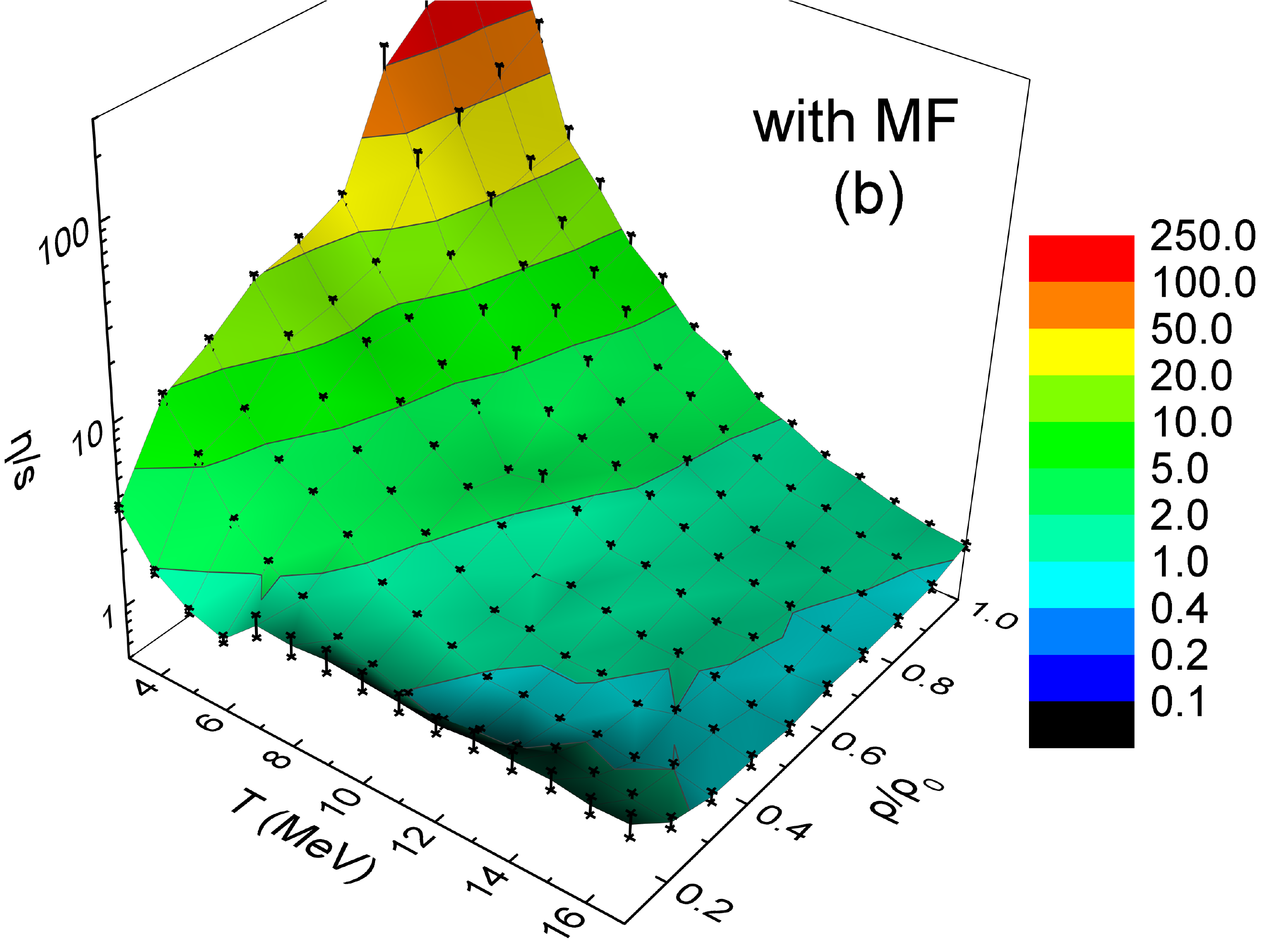}
\caption{\label{etas_3D} Specific shear viscosity $\eta/s$ in the $(T, \rho)$ plane from IBUU simulations in a box system without (a) and with (b) mean-field potential (MF).}
\end{figure}

Figure~\ref{etas_3D} provides a global picture of the specific shear viscosity in the $(T, \rho)$ plane with and without mean-field potential. The general feature that $\eta/s$ is large at higher densities and/or lower temperatures is seen in both cases. The distribution of $\eta/s$ in uniform systems without mean-field potential in the $(T, \rho)$ plane is seen to be flatter, while a concavity is seen in non-uniform systems with mean-field potential at lower densities. For a constant and isotropic nucleon-nucleon collision cross section $\sigma=40$ mb used in the present study, the value of $\eta/s$ is much larger than the KSS bound $\hbar/4\pi \approx 0.08$.
\\
\section{Summary and outlook}
\label{sec:summary}

Based on simulations in a box system with the periodic boundary condition using the IBUU transport model, we have studied the specific shear viscosity $\eta/s$ of nuclear matter at the average densities and temperatures around the spinodal region through the Green-Kubo method. The present study is based on previous efforts of the transport model evaluation project with well calibrated nucleon-nucleon collisions and mean-field evolutions. With the momentum-independent mean-field potential, which reproduces empirical nuclear matter properties and nuclear phase diagram, incorporated through the lattice Hamiltonian framework, we have generated dynamically stable and thermalized systems with nuclear clusters. By comparing results of the specific shear viscosity at different average densities and temperatures in uniform and non-uniform systems, we found that clusterizations may enhance the correlation of the energy-momentum tensor and the collision effect, thus reducing both $\eta$ and $\eta/s$. This leads to a minimum of $\eta/s$ as a function of temperature at lower average densities, while the minimum behavior disappears at $\rho > 0.3\rho_0$. Incorporating the Coulomb force reduces the clustering effect while the results remain qualitatively similar. The present study serves as a rigorous baseline calculation of $\eta/s$ with nuclear clusters, and helps to understand the relation between the shear viscosity and the nuclear phase diagram.

The study can be easily generalized to incorporate isospin degree of freedom as well as more realistic nucleon-nucleon collision cross sections. The Green-Kubo method can also be used to study other transport coefficients, e.g., the bulk viscosity. It is of great interest to study transport properties of isospin asymmetric nuclear matter in the mechanical and chemical instability region through the Green-Kubo method. Such studies are in progress.

\begin{acknowledgments}
JX is supported by the National Natural Science Foundation of China under Grant No. 11922514.
\end{acknowledgments}

\bibliography{box-viscosity}

\begin{thebibliography}{38}%
\makeatletter
\providecommand \@ifxundefined [1]{%
 \@ifx{#1\undefined}
}%
\providecommand \@ifnum [1]{%
 \ifnum #1\expandafter \@firstoftwo
 \else \expandafter \@secondoftwo
 \fi
}%
\providecommand \@ifx [1]{%
 \ifx #1\expandafter \@firstoftwo
 \else \expandafter \@secondoftwo
 \fi
}%
\providecommand \natexlab [1]{#1}%
\providecommand \enquote  [1]{``#1''}%
\providecommand \bibnamefont  [1]{#1}%
\providecommand \bibfnamefont [1]{#1}%
\providecommand \citenamefont [1]{#1}%
\providecommand \href@noop [0]{\@secondoftwo}%
\providecommand \href [0]{\begingroup \@sanitize@url \@href}%
\providecommand \@href[1]{\@@startlink{#1}\@@href}%
\providecommand \@@href[1]{\endgroup#1\@@endlink}%
\providecommand \@sanitize@url [0]{\catcode `\\12\catcode `\$12\catcode
  `\&12\catcode `\#12\catcode `\^12\catcode `\_12\catcode `\%12\relax}%
\providecommand \@@startlink[1]{}%
\providecommand \@@endlink[0]{}%
\providecommand \url  [0]{\begingroup\@sanitize@url \@url }%
\providecommand \@url [1]{\endgroup\@href {#1}{\urlprefix }}%
\providecommand \urlprefix  [0]{URL }%
\providecommand \Eprint [0]{\href }%
\providecommand \doibase [0]{http://dx.doi.org/}%
\providecommand \selectlanguage [0]{\@gobble}%
\providecommand \bibinfo  [0]{\@secondoftwo}%
\providecommand \bibfield  [0]{\@secondoftwo}%
\providecommand \translation [1]{[#1]}%
\providecommand \BibitemOpen [0]{}%
\providecommand \bibitemStop [0]{}%
\providecommand \bibitemNoStop [0]{.\EOS\space}%
\providecommand \EOS [0]{\spacefactor3000\relax}%
\providecommand \BibitemShut  [1]{\csname bibitem#1\endcsname}%
\let\auto@bib@innerbib\@empty
\bibitem [{\citenamefont {Peshier}\ and\ \citenamefont
  {Cassing}(2005)}]{Peshier:2005pp}%
  \BibitemOpen
  \bibfield  {author} {\bibinfo {author} {\bibfnamefont {A.}~\bibnamefont
  {Peshier}}\ and\ \bibinfo {author} {\bibfnamefont {W.}~\bibnamefont
  {Cassing}},\ }\bibfield  {title} {\enquote {\bibinfo {title} {{The Hot
  non-perturbative gluon plasma is an almost ideal colored liquid}},}\ }\href
  {\doibase 10.1103/PhysRevLett.94.172301} {\bibfield  {journal} {\bibinfo
  {journal} {Phys. Rev. Lett.}\ }\textbf {\bibinfo {volume} {94}},\ \bibinfo
  {pages} {172301} (\bibinfo {year} {2005})},\ \Eprint
  {http://arxiv.org/abs/hep-ph/0502138} {arXiv:hep-ph/0502138} \BibitemShut
  {NoStop}%
\bibitem [{\citenamefont {Majumder}\ \emph {et~al.}(2007)\citenamefont
  {Majumder}, \citenamefont {Muller},\ and\ \citenamefont
  {Wang}}]{Majumder:2007zh}%
  \BibitemOpen
  \bibfield  {author} {\bibinfo {author} {\bibfnamefont {Abhijit}\ \bibnamefont
  {Majumder}}, \bibinfo {author} {\bibfnamefont {Berndt}\ \bibnamefont
  {Muller}}, \ and\ \bibinfo {author} {\bibfnamefont {Xin-Nian}\ \bibnamefont
  {Wang}},\ }\bibfield  {title} {\enquote {\bibinfo {title} {{Small shear
  viscosity of a quark-gluon plasma implies strong jet quenching}},}\ }\href
  {\doibase 10.1103/PhysRevLett.99.192301} {\bibfield  {journal} {\bibinfo
  {journal} {Phys. Rev. Lett.}\ }\textbf {\bibinfo {volume} {99}},\ \bibinfo
  {pages} {192301} (\bibinfo {year} {2007})},\ \Eprint
  {http://arxiv.org/abs/hep-ph/0703082} {arXiv:hep-ph/0703082} \BibitemShut
  {NoStop}%
\bibitem [{\citenamefont {Song}\ \emph {et~al.}(2011)\citenamefont {Song},
  \citenamefont {Bass}, \citenamefont {Heinz}, \citenamefont {Hirano},\ and\
  \citenamefont {Shen}}]{Song:2010mg}%
  \BibitemOpen
  \bibfield  {author} {\bibinfo {author} {\bibfnamefont {Huichao}\ \bibnamefont
  {Song}}, \bibinfo {author} {\bibfnamefont {Steffen~A.}\ \bibnamefont {Bass}},
  \bibinfo {author} {\bibfnamefont {Ulrich}\ \bibnamefont {Heinz}}, \bibinfo
  {author} {\bibfnamefont {Tetsufumi}\ \bibnamefont {Hirano}}, \ and\ \bibinfo
  {author} {\bibfnamefont {Chun}\ \bibnamefont {Shen}},\ }\bibfield  {title}
  {\enquote {\bibinfo {title} {{200 A GeV Au+Au collisions serve a nearly
  perfect quark-gluon liquid}},}\ }\href {\doibase
  10.1103/PhysRevLett.106.192301} {\bibfield  {journal} {\bibinfo  {journal}
  {Phys. Rev. Lett.}\ }\textbf {\bibinfo {volume} {106}},\ \bibinfo {pages}
  {192301} (\bibinfo {year} {2011})},\ \bibinfo {note} {[Erratum:
  Phys.Rev.Lett. 109, 139904 (2012)]},\ \Eprint
  {http://arxiv.org/abs/1011.2783} {arXiv:1011.2783 [nucl-th]} \BibitemShut
  {NoStop}%
\bibitem [{\citenamefont {Schenke}\ \emph {et~al.}(2011)\citenamefont
  {Schenke}, \citenamefont {Jeon},\ and\ \citenamefont
  {Gale}}]{Schenke:2010rr}%
  \BibitemOpen
  \bibfield  {author} {\bibinfo {author} {\bibfnamefont {Bjorn}\ \bibnamefont
  {Schenke}}, \bibinfo {author} {\bibfnamefont {Sangyong}\ \bibnamefont
  {Jeon}}, \ and\ \bibinfo {author} {\bibfnamefont {Charles}\ \bibnamefont
  {Gale}},\ }\bibfield  {title} {\enquote {\bibinfo {title} {{Elliptic and
  triangular flow in event-by-event (3+1)D viscous hydrodynamics}},}\ }\href
  {\doibase 10.1103/PhysRevLett.106.042301} {\bibfield  {journal} {\bibinfo
  {journal} {Phys. Rev. Lett.}\ }\textbf {\bibinfo {volume} {106}},\ \bibinfo
  {pages} {042301} (\bibinfo {year} {2011})},\ \Eprint
  {http://arxiv.org/abs/1009.3244} {arXiv:1009.3244 [hep-ph]} \BibitemShut
  {NoStop}%
\bibitem [{\citenamefont {Bernhard}\ \emph {et~al.}(2019)\citenamefont
  {Bernhard}, \citenamefont {Moreland},\ and\ \citenamefont
  {Bass}}]{Bernhard:2019bmu}%
  \BibitemOpen
  \bibfield  {author} {\bibinfo {author} {\bibfnamefont {Jonah~E.}\
  \bibnamefont {Bernhard}}, \bibinfo {author} {\bibfnamefont {J.~Scott}\
  \bibnamefont {Moreland}}, \ and\ \bibinfo {author} {\bibfnamefont
  {Steffen~A.}\ \bibnamefont {Bass}},\ }\bibfield  {title} {\enquote {\bibinfo
  {title} {{Bayesian estimation of the specific shear and bulk viscosity of
  quark\textendash{}gluon plasma}},}\ }\href {\doibase
  10.1038/s41567-019-0611-8} {\bibfield  {journal} {\bibinfo  {journal} {Nature
  Phys.}\ }\textbf {\bibinfo {volume} {15}},\ \bibinfo {pages} {1113--1117}
  (\bibinfo {year} {2019})}\BibitemShut {NoStop}%
\bibitem [{\citenamefont {Parkkila}\ \emph {et~al.}(2022)\citenamefont
  {Parkkila}, \citenamefont {Onnerstad}, \citenamefont {Taghavi}, \citenamefont
  {Mordasini}, \citenamefont {Bilandzic}, \citenamefont {Virta},\ and\
  \citenamefont {Kim}}]{Parkkila:2021yha}%
  \BibitemOpen
  \bibfield  {author} {\bibinfo {author} {\bibfnamefont {J.~E.}\ \bibnamefont
  {Parkkila}}, \bibinfo {author} {\bibfnamefont {A.}~\bibnamefont {Onnerstad}},
  \bibinfo {author} {\bibfnamefont {S.~F.}\ \bibnamefont {Taghavi}}, \bibinfo
  {author} {\bibfnamefont {C.}~\bibnamefont {Mordasini}}, \bibinfo {author}
  {\bibfnamefont {A.}~\bibnamefont {Bilandzic}}, \bibinfo {author}
  {\bibfnamefont {M.}~\bibnamefont {Virta}}, \ and\ \bibinfo {author}
  {\bibfnamefont {D.~J.}\ \bibnamefont {Kim}},\ }\bibfield  {title} {\enquote
  {\bibinfo {title} {{New constraints for QCD matter from improved Bayesian
  parameter estimation in heavy-ion collisions at LHC}},}\ }\href {\doibase
  10.1016/j.physletb.2022.137485} {\bibfield  {journal} {\bibinfo  {journal}
  {Phys. Lett. B}\ }\textbf {\bibinfo {volume} {835}},\ \bibinfo {pages}
  {137485} (\bibinfo {year} {2022})},\ \Eprint
  {http://arxiv.org/abs/2111.08145} {arXiv:2111.08145 [hep-ph]} \BibitemShut
  {NoStop}%
\bibitem [{\citenamefont {Kovtun}\ \emph {et~al.}(2005)\citenamefont {Kovtun},
  \citenamefont {Son},\ and\ \citenamefont {Starinets}}]{Kovtun:2004de}%
  \BibitemOpen
  \bibfield  {author} {\bibinfo {author} {\bibfnamefont {P.}~\bibnamefont
  {Kovtun}}, \bibinfo {author} {\bibfnamefont {Dan~T.}\ \bibnamefont {Son}}, \
  and\ \bibinfo {author} {\bibfnamefont {Andrei~O.}\ \bibnamefont
  {Starinets}},\ }\bibfield  {title} {\enquote {\bibinfo {title} {{Viscosity in
  strongly interacting quantum field theories from black hole physics}},}\
  }\href {\doibase 10.1103/PhysRevLett.94.111601} {\bibfield  {journal}
  {\bibinfo  {journal} {Phys. Rev. Lett.}\ }\textbf {\bibinfo {volume} {94}},\
  \bibinfo {pages} {111601} (\bibinfo {year} {2005})},\ \Eprint
  {http://arxiv.org/abs/hep-th/0405231} {arXiv:hep-th/0405231} \BibitemShut
  {NoStop}%
\bibitem [{\citenamefont {Reichert}\ \emph {et~al.}(2021)\citenamefont
  {Reichert}, \citenamefont {Inghirami},\ and\ \citenamefont
  {Bleicher}}]{Reichert:2020oes}%
  \BibitemOpen
  \bibfield  {author} {\bibinfo {author} {\bibfnamefont {Tom}\ \bibnamefont
  {Reichert}}, \bibinfo {author} {\bibfnamefont {Gabriele}\ \bibnamefont
  {Inghirami}}, \ and\ \bibinfo {author} {\bibfnamefont {Marcus}\ \bibnamefont
  {Bleicher}},\ }\bibfield  {title} {\enquote {\bibinfo {title} {{A first
  estimate of $\eta/s$ in Au+Au reactions at $E_{lab}$ = 1.23 A GeV}},}\ }\href
  {\doibase 10.1016/j.physletb.2021.136285} {\bibfield  {journal} {\bibinfo
  {journal} {Phys. Lett. B}\ }\textbf {\bibinfo {volume} {817}},\ \bibinfo
  {pages} {136285} (\bibinfo {year} {2021})},\ \Eprint
  {http://arxiv.org/abs/2011.04546} {arXiv:2011.04546 [nucl-th]} \BibitemShut
  {NoStop}%
\bibitem [{\citenamefont {Csernai}\ \emph {et~al.}(2006)\citenamefont
  {Csernai}, \citenamefont {Kapusta},\ and\ \citenamefont
  {McLerran}}]{Csernai:2006zz}%
  \BibitemOpen
  \bibfield  {author} {\bibinfo {author} {\bibfnamefont {Laszlo~P.}\
  \bibnamefont {Csernai}}, \bibinfo {author} {\bibfnamefont {Joseph.~I.}\
  \bibnamefont {Kapusta}}, \ and\ \bibinfo {author} {\bibfnamefont {Larry~D.}\
  \bibnamefont {McLerran}},\ }\bibfield  {title} {\enquote {\bibinfo {title}
  {{On the Strongly-Interacting Low-Viscosity Matter Created in Relativistic
  Nuclear Collisions}},}\ }\href {\doibase 10.1103/PhysRevLett.97.152303}
  {\bibfield  {journal} {\bibinfo  {journal} {Phys. Rev. Lett.}\ }\textbf
  {\bibinfo {volume} {97}},\ \bibinfo {pages} {152303} (\bibinfo {year}
  {2006})},\ \Eprint {http://arxiv.org/abs/nucl-th/0604032}
  {arXiv:nucl-th/0604032} \BibitemShut {NoStop}%
\bibitem [{\citenamefont {Lacey}\ \emph {et~al.}(2007)\citenamefont {Lacey},
  \citenamefont {Ajitanand}, \citenamefont {Alexander}, \citenamefont {Chung},
  \citenamefont {Holzmann}, \citenamefont {Issah}, \citenamefont {Taranenko},
  \citenamefont {Danielewicz},\ and\ \citenamefont {Stoecker}}]{Lacey:2006bc}%
  \BibitemOpen
  \bibfield  {author} {\bibinfo {author} {\bibfnamefont {Roy~A.}\ \bibnamefont
  {Lacey}}, \bibinfo {author} {\bibfnamefont {N.~N.}\ \bibnamefont
  {Ajitanand}}, \bibinfo {author} {\bibfnamefont {J.~M.}\ \bibnamefont
  {Alexander}}, \bibinfo {author} {\bibfnamefont {P.}~\bibnamefont {Chung}},
  \bibinfo {author} {\bibfnamefont {W.~G.}\ \bibnamefont {Holzmann}}, \bibinfo
  {author} {\bibfnamefont {M.}~\bibnamefont {Issah}}, \bibinfo {author}
  {\bibfnamefont {A.}~\bibnamefont {Taranenko}}, \bibinfo {author}
  {\bibfnamefont {P.}~\bibnamefont {Danielewicz}}, \ and\ \bibinfo {author}
  {\bibfnamefont {Horst}\ \bibnamefont {Stoecker}},\ }\bibfield  {title}
  {\enquote {\bibinfo {title} {{Has the QCD Critical Point been Signaled by
  Observations at RHIC?}}}\ }\href {\doibase 10.1103/PhysRevLett.98.092301}
  {\bibfield  {journal} {\bibinfo  {journal} {Phys. Rev. Lett.}\ }\textbf
  {\bibinfo {volume} {98}},\ \bibinfo {pages} {092301} (\bibinfo {year}
  {2007})},\ \Eprint {http://arxiv.org/abs/nucl-ex/0609025}
  {arXiv:nucl-ex/0609025} \BibitemShut {NoStop}%
\bibitem [{\citenamefont {Chen}\ \emph {et~al.}(2007)\citenamefont {Chen},
  \citenamefont {Li}, \citenamefont {Liu},\ and\ \citenamefont
  {Nakano}}]{Chen:2007xe}%
  \BibitemOpen
  \bibfield  {author} {\bibinfo {author} {\bibfnamefont {Jiunn-Wei}\
  \bibnamefont {Chen}}, \bibinfo {author} {\bibfnamefont {Yen-Han}\
  \bibnamefont {Li}}, \bibinfo {author} {\bibfnamefont {Yen-Fu}\ \bibnamefont
  {Liu}}, \ and\ \bibinfo {author} {\bibfnamefont {Eiji}\ \bibnamefont
  {Nakano}},\ }\bibfield  {title} {\enquote {\bibinfo {title} {{QCD viscosity
  to entropy density ratio in the hadronic phase}},}\ }\href {\doibase
  10.1103/PhysRevD.76.114011} {\bibfield  {journal} {\bibinfo  {journal} {Phys.
  Rev. D}\ }\textbf {\bibinfo {volume} {76}},\ \bibinfo {pages} {114011}
  (\bibinfo {year} {2007})},\ \Eprint {http://arxiv.org/abs/hep-ph/0703230}
  {arXiv:hep-ph/0703230} \BibitemShut {NoStop}%
\bibitem [{\citenamefont {Pal}(2010)}]{Pal:2010sj}%
  \BibitemOpen
  \bibfield  {author} {\bibinfo {author} {\bibfnamefont {Subrata}\ \bibnamefont
  {Pal}},\ }\bibfield  {title} {\enquote {\bibinfo {title} {{Shear viscosity to
  entropy density ratio in nuclear multifragmentation}},}\ }\href {\doibase
  10.1103/PhysRevC.81.051601} {\bibfield  {journal} {\bibinfo  {journal} {Phys.
  Rev. C}\ }\textbf {\bibinfo {volume} {81}},\ \bibinfo {pages} {051601}
  (\bibinfo {year} {2010})},\ \Eprint {http://arxiv.org/abs/1005.0227}
  {arXiv:1005.0227 [nucl-th]} \BibitemShut {NoStop}%
\bibitem [{\citenamefont {Xu}\ \emph {et~al.}(2013)\citenamefont {Xu},
  \citenamefont {Chen}, \citenamefont {Ko}, \citenamefont {Li},\ and\
  \citenamefont {Ma}}]{Xu:2013nwa}%
  \BibitemOpen
  \bibfield  {author} {\bibinfo {author} {\bibfnamefont {Jun}\ \bibnamefont
  {Xu}}, \bibinfo {author} {\bibfnamefont {Lie-Wen}\ \bibnamefont {Chen}},
  \bibinfo {author} {\bibfnamefont {Che~Ming}\ \bibnamefont {Ko}}, \bibinfo
  {author} {\bibfnamefont {Bao-An}\ \bibnamefont {Li}}, \ and\ \bibinfo
  {author} {\bibfnamefont {Yu-Gang}\ \bibnamefont {Ma}},\ }\bibfield  {title}
  {\enquote {\bibinfo {title} {{Shear viscosity of neutron-rich nucleonic
  matter near its liquid-gas phase transition}},}\ }\href {\doibase
  10.1016/j.physletb.2013.10.051} {\bibfield  {journal} {\bibinfo  {journal}
  {Phys. Lett. B}\ }\textbf {\bibinfo {volume} {727}},\ \bibinfo {pages}
  {244--248} (\bibinfo {year} {2013})},\ \Eprint
  {http://arxiv.org/abs/1306.5361} {arXiv:1306.5361 [nucl-th]} \BibitemShut
  {NoStop}%
\bibitem [{\citenamefont {Xu}(2015)}]{Xu:2015lna}%
  \BibitemOpen
  \bibfield  {author} {\bibinfo {author} {\bibfnamefont {Jun}\ \bibnamefont
  {Xu}},\ }\bibfield  {title} {\enquote {\bibinfo {title} {{Isospin splitting
  of nucleon effective mass and shear viscosity of nuclear matter}},}\ }\href
  {\doibase 10.1103/PhysRevC.91.037601} {\bibfield  {journal} {\bibinfo
  {journal} {Phys. Rev. C}\ }\textbf {\bibinfo {volume} {91}},\ \bibinfo
  {pages} {037601} (\bibinfo {year} {2015})},\ \Eprint
  {http://arxiv.org/abs/1502.02335} {arXiv:1502.02335 [nucl-th]} \BibitemShut
  {NoStop}%
\bibitem [{\citenamefont {Deng}\ \emph {et~al.}(2022)\citenamefont {Deng},
  \citenamefont {Danielewicz}, \citenamefont {Ma}, \citenamefont {Lin},\ and\
  \citenamefont {Zhang}}]{PhysRevC.105.064613}%
  \BibitemOpen
  \bibfield  {author} {\bibinfo {author} {\bibfnamefont {X.~G.}\ \bibnamefont
  {Deng}}, \bibinfo {author} {\bibfnamefont {P.}~\bibnamefont {Danielewicz}},
  \bibinfo {author} {\bibfnamefont {Y.~G.}\ \bibnamefont {Ma}}, \bibinfo
  {author} {\bibfnamefont {H.}~\bibnamefont {Lin}}, \ and\ \bibinfo {author}
  {\bibfnamefont {Y.~X.}\ \bibnamefont {Zhang}},\ }\bibfield  {title} {\enquote
  {\bibinfo {title} {Impact of fragment formation on shear viscosity in the
  nuclear liquid-gas phase transition region},}\ }\href {\doibase
  10.1103/PhysRevC.105.064613} {\bibfield  {journal} {\bibinfo  {journal}
  {Phys. Rev. C}\ }\textbf {\bibinfo {volume} {105}},\ \bibinfo {pages}
  {064613} (\bibinfo {year} {2022})}\BibitemShut {NoStop}%
\bibitem [{\citenamefont {Ghosh}\ \emph {et~al.}(2015)\citenamefont {Ghosh},
  \citenamefont {Raha}, \citenamefont {Ray}, \citenamefont {Saha},\ and\
  \citenamefont {Upadhaya}}]{Ghosh:2014vja}%
  \BibitemOpen
  \bibfield  {author} {\bibinfo {author} {\bibfnamefont {Sanjay~K.}\
  \bibnamefont {Ghosh}}, \bibinfo {author} {\bibfnamefont {Sibaji}\
  \bibnamefont {Raha}}, \bibinfo {author} {\bibfnamefont {Rajarshi}\
  \bibnamefont {Ray}}, \bibinfo {author} {\bibfnamefont {Kinkar}\ \bibnamefont
  {Saha}}, \ and\ \bibinfo {author} {\bibfnamefont {Sudipa}\ \bibnamefont
  {Upadhaya}},\ }\bibfield  {title} {\enquote {\bibinfo {title} {{Shear
  viscosity and phase diagram from
  Polyakov\textendash{}Nambu\textendash{}Jona-Lasinio model}},}\ }\href
  {\doibase 10.1103/PhysRevD.91.054005} {\bibfield  {journal} {\bibinfo
  {journal} {Phys. Rev. D}\ }\textbf {\bibinfo {volume} {91}},\ \bibinfo
  {pages} {054005} (\bibinfo {year} {2015})},\ \Eprint
  {http://arxiv.org/abs/1411.2765} {arXiv:1411.2765 [hep-ph]} \BibitemShut
  {NoStop}%
\bibitem [{\citenamefont {Grefa}\ \emph {et~al.}(2022)\citenamefont {Grefa},
  \citenamefont {Hippert}, \citenamefont {Noronha}, \citenamefont
  {Noronha-Hostler}, \citenamefont {Portillo}, \citenamefont {Ratti},\ and\
  \citenamefont {Rougemont}}]{Grefa:2022sav}%
  \BibitemOpen
  \bibfield  {author} {\bibinfo {author} {\bibfnamefont {Joaquin}\ \bibnamefont
  {Grefa}}, \bibinfo {author} {\bibfnamefont {Mauricio}\ \bibnamefont
  {Hippert}}, \bibinfo {author} {\bibfnamefont {Jorge}\ \bibnamefont
  {Noronha}}, \bibinfo {author} {\bibfnamefont {Jacquelyn}\ \bibnamefont
  {Noronha-Hostler}}, \bibinfo {author} {\bibfnamefont {Israel}\ \bibnamefont
  {Portillo}}, \bibinfo {author} {\bibfnamefont {Claudia}\ \bibnamefont
  {Ratti}}, \ and\ \bibinfo {author} {\bibfnamefont {Romulo}\ \bibnamefont
  {Rougemont}},\ }\bibfield  {title} {\enquote {\bibinfo {title} {{Transport
  coefficients of the quark-gluon plasma at the critical point and across the
  first-order line}},}\ }\href {\doibase 10.1103/PhysRevD.106.034024}
  {\bibfield  {journal} {\bibinfo  {journal} {Phys. Rev. D}\ }\textbf {\bibinfo
  {volume} {106}},\ \bibinfo {pages} {034024} (\bibinfo {year} {2022})},\
  \Eprint {http://arxiv.org/abs/2203.00139} {arXiv:2203.00139 [nucl-th]}
  \BibitemShut {NoStop}%
\bibitem [{\citenamefont {Muronga}(2004)}]{Muronga:2003tb}%
  \BibitemOpen
  \bibfield  {author} {\bibinfo {author} {\bibfnamefont {Azwinndini}\
  \bibnamefont {Muronga}},\ }\bibfield  {title} {\enquote {\bibinfo {title}
  {{Shear viscosity coefficient from microscopic models}},}\ }\href {\doibase
  10.1103/PhysRevC.69.044901} {\bibfield  {journal} {\bibinfo  {journal} {Phys.
  Rev. C}\ }\textbf {\bibinfo {volume} {69}},\ \bibinfo {pages} {044901}
  (\bibinfo {year} {2004})},\ \Eprint {http://arxiv.org/abs/nucl-th/0309056}
  {arXiv:nucl-th/0309056} \BibitemShut {NoStop}%
\bibitem [{\citenamefont {Chen}\ and\ \citenamefont
  {Nakano}(2007)}]{Chen:2006iga}%
  \BibitemOpen
  \bibfield  {author} {\bibinfo {author} {\bibfnamefont {Jiunn-Wei}\
  \bibnamefont {Chen}}\ and\ \bibinfo {author} {\bibfnamefont {Eiji}\
  \bibnamefont {Nakano}},\ }\bibfield  {title} {\enquote {\bibinfo {title}
  {{Shear viscosity to entropy density ratio of QCD below the deconfinement
  temperature}},}\ }\href {\doibase 10.1016/j.physletb.2007.02.026} {\bibfield
  {journal} {\bibinfo  {journal} {Phys. Lett. B}\ }\textbf {\bibinfo {volume}
  {647}},\ \bibinfo {pages} {371--375} (\bibinfo {year} {2007})},\ \Eprint
  {http://arxiv.org/abs/hep-ph/0604138} {arXiv:hep-ph/0604138} \BibitemShut
  {NoStop}%
\bibitem [{\citenamefont {Demir}\ and\ \citenamefont
  {Bass}(2009)}]{Demir:2008tr}%
  \BibitemOpen
  \bibfield  {author} {\bibinfo {author} {\bibfnamefont {Nasser}\ \bibnamefont
  {Demir}}\ and\ \bibinfo {author} {\bibfnamefont {Steffen~A.}\ \bibnamefont
  {Bass}},\ }\bibfield  {title} {\enquote {\bibinfo {title} {{Shear-Viscosity
  to Entropy-Density Ratio of a Relativistic Hadron Gas}},}\ }\href {\doibase
  10.1103/PhysRevLett.102.172302} {\bibfield  {journal} {\bibinfo  {journal}
  {Phys. Rev. Lett.}\ }\textbf {\bibinfo {volume} {102}},\ \bibinfo {pages}
  {172302} (\bibinfo {year} {2009})},\ \Eprint {http://arxiv.org/abs/0812.2422}
  {arXiv:0812.2422 [nucl-th]} \BibitemShut {NoStop}%
\bibitem [{\citenamefont {Rose}\ \emph {et~al.}(2018)\citenamefont {Rose},
  \citenamefont {Torres-Rincon}, \citenamefont {Sch\"afer}, \citenamefont
  {Oliinychenko},\ and\ \citenamefont {Petersen}}]{Rose:2017bjz}%
  \BibitemOpen
  \bibfield  {author} {\bibinfo {author} {\bibfnamefont {J.~B.}\ \bibnamefont
  {Rose}}, \bibinfo {author} {\bibfnamefont {J.~M.}\ \bibnamefont
  {Torres-Rincon}}, \bibinfo {author} {\bibfnamefont {A.}~\bibnamefont
  {Sch\"afer}}, \bibinfo {author} {\bibfnamefont {D.~R.}\ \bibnamefont
  {Oliinychenko}}, \ and\ \bibinfo {author} {\bibfnamefont {H.}~\bibnamefont
  {Petersen}},\ }\bibfield  {title} {\enquote {\bibinfo {title} {{Shear
  viscosity of a hadron gas and influence of resonance lifetimes on relaxation
  time}},}\ }\href {\doibase 10.1103/PhysRevC.97.055204} {\bibfield  {journal}
  {\bibinfo  {journal} {Phys. Rev. C}\ }\textbf {\bibinfo {volume} {97}},\
  \bibinfo {pages} {055204} (\bibinfo {year} {2018})},\ \Eprint
  {http://arxiv.org/abs/1709.03826} {arXiv:1709.03826 [nucl-th]} \BibitemShut
  {NoStop}%
\bibitem [{\citenamefont {Danielewicz}(1984)}]{Danielewicz:1984kt}%
  \BibitemOpen
  \bibfield  {author} {\bibinfo {author} {\bibfnamefont {P.}~\bibnamefont
  {Danielewicz}},\ }\bibfield  {title} {\enquote {\bibinfo {title} {{TRANSPORT
  PROPERTIES OF EXCITED NUCLEAR MATTER AND THE SHOCK WAVE PROFILE}},}\ }\href
  {\doibase 10.1016/0370-2693(84)91010-4} {\bibfield  {journal} {\bibinfo
  {journal} {Phys. Lett. B}\ }\textbf {\bibinfo {volume} {146}},\ \bibinfo
  {pages} {168--175} (\bibinfo {year} {1984})}\BibitemShut {NoStop}%
\bibitem [{\citenamefont {Shi}\ and\ \citenamefont
  {Danielewicz}(2003)}]{Shi:2003np}%
  \BibitemOpen
  \bibfield  {author} {\bibinfo {author} {\bibfnamefont {L.}~\bibnamefont
  {Shi}}\ and\ \bibinfo {author} {\bibfnamefont {P.}~\bibnamefont
  {Danielewicz}},\ }\bibfield  {title} {\enquote {\bibinfo {title} {{Nuclear
  isospin diffusivity}},}\ }\href {\doibase 10.1103/PhysRevC.68.064604}
  {\bibfield  {journal} {\bibinfo  {journal} {Phys. Rev. C}\ }\textbf {\bibinfo
  {volume} {68}},\ \bibinfo {pages} {064604} (\bibinfo {year} {2003})},\
  \Eprint {http://arxiv.org/abs/nucl-th/0304030} {arXiv:nucl-th/0304030}
  \BibitemShut {NoStop}%
\bibitem [{\citenamefont {Kubo}(1966)}]{Kubo_1966}%
  \BibitemOpen
  \bibfield  {author} {\bibinfo {author} {\bibfnamefont {R.}~\bibnamefont
  {Kubo}},\ }\bibfield  {title} {\enquote {\bibinfo {title} {The
  fluctuation-dissipation theorem},}\ }\href {\doibase
  10.1088/0034-4885/29/1/306} {\bibfield  {journal} {\bibinfo  {journal}
  {Reports on Progress in Physics}\ }\textbf {\bibinfo {volume} {29}},\
  \bibinfo {pages} {255--284} (\bibinfo {year} {1966})}\BibitemShut {NoStop}%
\bibitem [{\citenamefont {Plumari}\ \emph {et~al.}(2012)\citenamefont
  {Plumari}, \citenamefont {Puglisi}, \citenamefont {Scardina},\ and\
  \citenamefont {Greco}}]{Plumari:2012ep}%
  \BibitemOpen
  \bibfield  {author} {\bibinfo {author} {\bibfnamefont {S.}~\bibnamefont
  {Plumari}}, \bibinfo {author} {\bibfnamefont {A.}~\bibnamefont {Puglisi}},
  \bibinfo {author} {\bibfnamefont {F.}~\bibnamefont {Scardina}}, \ and\
  \bibinfo {author} {\bibfnamefont {V.}~\bibnamefont {Greco}},\ }\bibfield
  {title} {\enquote {\bibinfo {title} {{Shear Viscosity of a strongly
  interacting system: Green-Kubo vs. Chapman-Enskog and Relaxation Time
  Approximation}},}\ }\href {\doibase 10.1103/PhysRevC.86.054902} {\bibfield
  {journal} {\bibinfo  {journal} {Phys. Rev. C}\ }\textbf {\bibinfo {volume}
  {86}},\ \bibinfo {pages} {054902} (\bibinfo {year} {2012})},\ \Eprint
  {http://arxiv.org/abs/1208.0481} {arXiv:1208.0481 [nucl-th]} \BibitemShut
  {NoStop}%
\bibitem [{\citenamefont {Xu}(2019)}]{Xu:2019hqg}%
  \BibitemOpen
  \bibfield  {author} {\bibinfo {author} {\bibfnamefont {Jun}\ \bibnamefont
  {Xu}},\ }\bibfield  {title} {\enquote {\bibinfo {title} {{Transport
  approaches for the description of intermediate-energy heavy-ion
  collisions}},}\ }\href {\doibase 10.1016/j.ppnp.2019.02.009} {\bibfield
  {journal} {\bibinfo  {journal} {Prog. Part. Nucl. Phys.}\ }\textbf {\bibinfo
  {volume} {106}},\ \bibinfo {pages} {312--359} (\bibinfo {year} {2019})},\
  \Eprint {http://arxiv.org/abs/1904.00131} {arXiv:1904.00131 [nucl-th]}
  \BibitemShut {NoStop}%
\bibitem [{\citenamefont {Motornenko}\ \emph {et~al.}(2018)\citenamefont
  {Motornenko}, \citenamefont {Bravina}, \citenamefont {Gorenstein},
  \citenamefont {Magner},\ and\ \citenamefont {Zabrodin}}]{Motornenko:2017hob}%
  \BibitemOpen
  \bibfield  {author} {\bibinfo {author} {\bibfnamefont {A.}~\bibnamefont
  {Motornenko}}, \bibinfo {author} {\bibfnamefont {L.}~\bibnamefont {Bravina}},
  \bibinfo {author} {\bibfnamefont {M.~I.}\ \bibnamefont {Gorenstein}},
  \bibinfo {author} {\bibfnamefont {A.~G.}\ \bibnamefont {Magner}}, \ and\
  \bibinfo {author} {\bibfnamefont {E.}~\bibnamefont {Zabrodin}},\ }\bibfield
  {title} {\enquote {\bibinfo {title} {{Nucleon matter equation of state,
  particle number fluctuations, and shear viscosity within UrQMD box
  calculations}},}\ }\href {\doibase 10.1088/1361-6471/aaa78a} {\bibfield
  {journal} {\bibinfo  {journal} {J. Phys. G}\ }\textbf {\bibinfo {volume}
  {45}},\ \bibinfo {pages} {035101} (\bibinfo {year} {2018})},\ \Eprint
  {http://arxiv.org/abs/1710.09276} {arXiv:1710.09276 [nucl-th]} \BibitemShut
  {NoStop}%
\bibitem [{\citenamefont {Deng}\ \emph {et~al.}(2021)\citenamefont {Deng},
  \citenamefont {Ma},\ and\ \citenamefont {Zhang}}]{Deng:2021rpq}%
  \BibitemOpen
  \bibfield  {author} {\bibinfo {author} {\bibfnamefont {X.~G.}\ \bibnamefont
  {Deng}}, \bibinfo {author} {\bibfnamefont {Y.~G.}\ \bibnamefont {Ma}}, \ and\
  \bibinfo {author} {\bibfnamefont {Y.~X.}\ \bibnamefont {Zhang}},\ }\bibfield
  {title} {\enquote {\bibinfo {title} {{Green\textendash{}Kubo formula for
  Boltzmann and Fermi\textendash{}Dirac statistics}},}\ }\href {\doibase
  10.1140/epja/s10050-021-00550-4} {\bibfield  {journal} {\bibinfo  {journal}
  {Eur. Phys. J. A}\ }\textbf {\bibinfo {volume} {57}},\ \bibinfo {pages} {242}
  (\bibinfo {year} {2021})},\ \Eprint {http://arxiv.org/abs/2109.09631}
  {arXiv:2109.09631 [nucl-th]} \BibitemShut {NoStop}%
\bibitem [{\citenamefont {Zhang}\ \emph {et~al.}(2018)\citenamefont {Zhang}
  \emph {et~al.}}]{Zhang:2017esm}%
  \BibitemOpen
  \bibfield  {author} {\bibinfo {author} {\bibfnamefont {Ying-Xun}\
  \bibnamefont {Zhang}} \emph {et~al.},\ }\bibfield  {title} {\enquote
  {\bibinfo {title} {{Comparison of heavy-ion transport simulations: Collision
  integral in a box}},}\ }\href {\doibase 10.1103/PhysRevC.97.034625}
  {\bibfield  {journal} {\bibinfo  {journal} {Phys. Rev. C}\ }\textbf {\bibinfo
  {volume} {97}},\ \bibinfo {pages} {034625} (\bibinfo {year} {2018})},\
  \Eprint {http://arxiv.org/abs/1711.05950} {arXiv:1711.05950 [nucl-th]}
  \BibitemShut {NoStop}%
\bibitem [{\citenamefont {Colonna}\ \emph {et~al.}(2021)\citenamefont {Colonna}
  \emph {et~al.}}]{Colonna:2021xuh}%
  \BibitemOpen
  \bibfield  {author} {\bibinfo {author} {\bibfnamefont {Maria}\ \bibnamefont
  {Colonna}} \emph {et~al.},\ }\bibfield  {title} {\enquote {\bibinfo {title}
  {{Comparison of heavy-ion transport simulations: Mean-field dynamics in a
  box}},}\ }\href {\doibase 10.1103/PhysRevC.104.024603} {\bibfield  {journal}
  {\bibinfo  {journal} {Phys. Rev. C}\ }\textbf {\bibinfo {volume} {104}},\
  \bibinfo {pages} {024603} (\bibinfo {year} {2021})},\ \Eprint
  {http://arxiv.org/abs/2106.12287} {arXiv:2106.12287 [nucl-th]} \BibitemShut
  {NoStop}%
\bibitem [{\citenamefont {Bertsch}\ and\ \citenamefont
  {Das~Gupta}(1988)}]{Bertsch:1988ik}%
  \BibitemOpen
  \bibfield  {author} {\bibinfo {author} {\bibfnamefont {G.~F.}\ \bibnamefont
  {Bertsch}}\ and\ \bibinfo {author} {\bibfnamefont {S.}~\bibnamefont
  {Das~Gupta}},\ }\bibfield  {title} {\enquote {\bibinfo {title} {{A Guide to
  microscopic models for intermediate-energy heavy ion collisions}},}\ }\href
  {\doibase 10.1016/0370-1573(88)90170-6} {\bibfield  {journal} {\bibinfo
  {journal} {Phys. Rept.}\ }\textbf {\bibinfo {volume} {160}},\ \bibinfo
  {pages} {189--233} (\bibinfo {year} {1988})}\BibitemShut {NoStop}%
\bibitem [{\citenamefont {Lenk}\ and\ \citenamefont
  {Pandharipande}(1989)}]{Lenk:1989zz}%
  \BibitemOpen
  \bibfield  {author} {\bibinfo {author} {\bibfnamefont {R.~J.}\ \bibnamefont
  {Lenk}}\ and\ \bibinfo {author} {\bibfnamefont {V.~R.}\ \bibnamefont
  {Pandharipande}},\ }\bibfield  {title} {\enquote {\bibinfo {title} {{Nuclear
  mean field dynamics in the lattice Hamiltonian Vlasov method}},}\ }\href
  {\doibase 10.1103/PhysRevC.39.2242} {\bibfield  {journal} {\bibinfo
  {journal} {Phys. Rev. C}\ }\textbf {\bibinfo {volume} {39}},\ \bibinfo
  {pages} {2242--2249} (\bibinfo {year} {1989})}\BibitemShut {NoStop}%
\bibitem [{\citenamefont {Xu}\ \emph {et~al.}(2008)\citenamefont {Xu},
  \citenamefont {Chen}, \citenamefont {Li},\ and\ \citenamefont
  {Ma}}]{Xu:2007eq}%
  \BibitemOpen
  \bibfield  {author} {\bibinfo {author} {\bibfnamefont {Jun}\ \bibnamefont
  {Xu}}, \bibinfo {author} {\bibfnamefont {Lie-Wen}\ \bibnamefont {Chen}},
  \bibinfo {author} {\bibfnamefont {Bao-An}\ \bibnamefont {Li}}, \ and\
  \bibinfo {author} {\bibfnamefont {Hong-Ru}\ \bibnamefont {Ma}},\ }\bibfield
  {title} {\enquote {\bibinfo {title} {{Effects of isospin and momentum
  dependent interactions on thermal properties of asymmetric nuclear
  matter}},}\ }\href {\doibase 10.1103/PhysRevC.77.014302} {\bibfield
  {journal} {\bibinfo  {journal} {Phys. Rev. C}\ }\textbf {\bibinfo {volume}
  {77}},\ \bibinfo {pages} {014302} (\bibinfo {year} {2008})},\ \Eprint
  {http://arxiv.org/abs/0710.5409} {arXiv:0710.5409 [nucl-th]} \BibitemShut
  {NoStop}%
\bibitem [{\citenamefont {Muller}\ and\ \citenamefont
  {Serot}(1995)}]{Muller:1995ji}%
  \BibitemOpen
  \bibfield  {author} {\bibinfo {author} {\bibfnamefont {Horst}\ \bibnamefont
  {Muller}}\ and\ \bibinfo {author} {\bibfnamefont {Brian~D.}\ \bibnamefont
  {Serot}},\ }\bibfield  {title} {\enquote {\bibinfo {title} {{Phase
  transitions in warm, asymmetric nuclear matter}},}\ }\href {\doibase
  10.1103/PhysRevC.52.2072} {\bibfield  {journal} {\bibinfo  {journal} {Phys.
  Rev. C}\ }\textbf {\bibinfo {volume} {52}},\ \bibinfo {pages} {2072--2091}
  (\bibinfo {year} {1995})},\ \Eprint {http://arxiv.org/abs/nucl-th/9505013}
  {arXiv:nucl-th/9505013} \BibitemShut {NoStop}%
\bibitem [{\citenamefont {Chomaz}\ \emph {et~al.}(2004)\citenamefont {Chomaz},
  \citenamefont {Colonna},\ and\ \citenamefont {Randrup}}]{Chomaz:2003dz}%
  \BibitemOpen
  \bibfield  {author} {\bibinfo {author} {\bibfnamefont {Philipe}\ \bibnamefont
  {Chomaz}}, \bibinfo {author} {\bibfnamefont {Maria}\ \bibnamefont {Colonna}},
  \ and\ \bibinfo {author} {\bibfnamefont {Jorgen}\ \bibnamefont {Randrup}},\
  }\bibfield  {title} {\enquote {\bibinfo {title} {{Nuclear spinodal
  fragmentation}},}\ }\href {\doibase 10.1016/j.physrep.2003.09.006} {\bibfield
   {journal} {\bibinfo  {journal} {Phys. Rept.}\ }\textbf {\bibinfo {volume}
  {389}},\ \bibinfo {pages} {263--440} (\bibinfo {year} {2004})}\BibitemShut
  {NoStop}%
\bibitem [{\citenamefont {Hosoya}\ \emph {et~al.}(1984)\citenamefont {Hosoya},
  \citenamefont {Sakagami},\ and\ \citenamefont {Takao}}]{Hosoya:1983id}%
  \BibitemOpen
  \bibfield  {author} {\bibinfo {author} {\bibfnamefont {Akio}\ \bibnamefont
  {Hosoya}}, \bibinfo {author} {\bibfnamefont {Masa-aki}\ \bibnamefont
  {Sakagami}}, \ and\ \bibinfo {author} {\bibfnamefont {Masaru}\ \bibnamefont
  {Takao}},\ }\bibfield  {title} {\enquote {\bibinfo {title} {{Nonequilibrium
  Thermodynamics in Field Theory: Transport Coefficients}},}\ }\href {\doibase
  10.1016/0003-4916(84)90144-1} {\bibfield  {journal} {\bibinfo  {journal}
  {Annals Phys.}\ }\textbf {\bibinfo {volume} {154}},\ \bibinfo {pages} {229}
  (\bibinfo {year} {1984})}\BibitemShut {NoStop}%
\bibitem [{\citenamefont {Paech}\ and\ \citenamefont
  {Pratt}(2006)}]{Paech:2006st}%
  \BibitemOpen
  \bibfield  {author} {\bibinfo {author} {\bibfnamefont {Kerstin}\ \bibnamefont
  {Paech}}\ and\ \bibinfo {author} {\bibfnamefont {Scott}\ \bibnamefont
  {Pratt}},\ }\bibfield  {title} {\enquote {\bibinfo {title} {{Origins of bulk
  viscosity in relativistic heavy ion collisions}},}\ }\href {\doibase
  10.1103/PhysRevC.74.014901} {\bibfield  {journal} {\bibinfo  {journal} {Phys.
  Rev. C}\ }\textbf {\bibinfo {volume} {74}},\ \bibinfo {pages} {014901}
  (\bibinfo {year} {2006})},\ \bibinfo {note} {[Erratum: Phys.Rev.C 93, 059902
  (2016)]},\ \Eprint {http://arxiv.org/abs/nucl-th/0604008}
  {arXiv:nucl-th/0604008} \BibitemShut {NoStop}%
\bibitem [{\citenamefont {Burgio}\ \emph {et~al.}(1992)\citenamefont {Burgio},
  \citenamefont {Chomaz},\ and\ \citenamefont {Randrup}}]{Burgio:1991ej}%
  \BibitemOpen
  \bibfield  {author} {\bibinfo {author} {\bibfnamefont {G.~F.}\ \bibnamefont
  {Burgio}}, \bibinfo {author} {\bibfnamefont {P.}~\bibnamefont {Chomaz}}, \
  and\ \bibinfo {author} {\bibfnamefont {J.}~\bibnamefont {Randrup}},\
  }\bibfield  {title} {\enquote {\bibinfo {title} {{Dynamical clusterization in
  presence of instabilities}},}\ }\href {\doibase 10.1103/PhysRevLett.69.885}
  {\bibfield  {journal} {\bibinfo  {journal} {Phys. Rev. Lett.}\ }\textbf
  {\bibinfo {volume} {69}},\ \bibinfo {pages} {885--888} (\bibinfo {year}
  {1992})}\BibitemShut {NoStop}%
\end{thebibliography}%
\end{document}